# Theory of Plasma-Cascade Instability

Vladimir N. Litvinenko*[1,2] and Gang Wang[2,1]

[1] Department of Physics and Astronomy, Stony Brook University, Stony Brook, NY
[2] Collider-Accelerator Department, Brookhaven National Laboratory, Upton, NY

*Abstract.* In this paper we present the theory of a novel micro-bunching instability occurring in charged particle beams propagating along a straight trajectory: a Plasma-Cascade Instability (PCI). This instability was confirmed by 3D numerical simulations [1] and observed experimentally [2]. It can by driven by variation of beam's density and/or particle's mobility. The PCI can strongly amplify noise in the beam and drastically reduce its quality. Conversely, such instability can drive novel high-power sources of radiation or can be used as a broadband amplifier.

## I.   Introduction

High brightness intense charged particle beams are central for high luminosity hadron colliders [3-7], X-ray free-electron-lasers (FEL) [8-22] as well as for future hadron beam coolers [23-24], X-ray FEL oscillators [26-29], and plasma-wake-field accelerators with TV/m accelerating gradients [30-38]. Preservation of the beam quality during generation, acceleration, transportation and compression is important for attaining the desirable properties of the beam. On the other hand, instabilities in the beams can be deliberately built-in to attain specific results. The most known application is the FEL instability used for generating coherent radiation from THz to X-rays [39-43]. Less known applications are Coherent electron Cooling (CeC) of hadron beams [44-49] or generation of broad-band high power radiation by micro-bunched beams [50-52].

The PCI is the micro-bunching instability occurs in a beam propagating along a straight line. It differs from the well-known and well-studied conventional micro-bunching instabilities [53-72] requiring bending of the beam trajectory[1].

Theory of the PCI is important both for understanding of the nature of this instability and for predicting condition when PCI can occur. We will start from discussions of a simple model for such instability and continue with a rigorous derivation of the integral equation describing the PCI in 3D case. We will continue with reducing the integral equation to a second order ordinary differential equation for two specific cases of momenta distribution. We conclude with deriving equation for longitudinal PCI and discussions of our PCI theory in the framework of other instabilities.

## II.   Simple model of Plasma-Cascade Instability

Let's consider an evolution of perturbations in a charged homogeneous beam using the co-moving frame of reference, where motion of particles can be considered non-relativistic. It is well known that small density perturbations $|\delta n|/n_o \ll 1$ in a cold homogeneous beam (e.g., non-neutral plasma) will undergo oscillations with plasma frequency [73-74],

$$\ddot{\tilde{n}} + \omega_p^2 \tilde{n} = 0; \quad \omega_p = c\sqrt{4\pi n_r r_c} \tag{1}$$

where $n_r$ is the particles density in the rest frame of particles, $c$ is the speed of light, $r_c = e^2/mc^2$ is the classical radius, $e$ is charge and $m$ is its rest mass of the particles [2]. Equation (1) has a trivial oscillatory solution for an arbitrary infinitesimal $\delta n(\vec{r})$:

$$\tilde{n} = \delta n(\vec{r}) \cdot \cos(\omega_p t + \varphi), \tag{2}$$

with constants of motion $\delta n, \varphi$ determined by initial conditions. During this stable oscillation after each quarter of the plasma period the electrostatic energy (e.g., that of the density perturbation) is transferred to the kinetic energy (e.g., velocity perturbation), and vice versa. In plasma with finite temperature this stable oscillation will eventually decay [73-74].

---

[1] For example, in a magnetic chicane or in an arc of an accelerator.

Situation does change if the density of the beam, and the corresponding plasma frequency, vary in time. The oscillator equation with time-dependent frequency

$$\ddot{\tilde{n}} + \omega_p^2(t)\tilde{n} = 0, \tag{3}$$

may result in unstable oscillations with exponentially growing amplitude. In general, Eq. (3) does not have an analytical solution. But its solution can be represented by a linear symplectic transformation [75-78], which can be formally expressed in the form of ordered matrix exponent:

$$\begin{bmatrix} \tilde{n}(t_2) \\ \dot{\tilde{n}}(t_2) \end{bmatrix} = \mathbf{M}(t_1|t_2) \begin{bmatrix} \tilde{n}(t_1) \\ \dot{\tilde{n}}(t_1) \end{bmatrix}; \mathbf{M}(t_1|t_2) = \underset{ordered}{exp}\left[\int_{t_1}^{t_2} \mathbf{D}(t)dt\right]; \mathbf{D}(t) = \begin{bmatrix} 0 & 1 \\ -\omega_p^2(t) & 0 \end{bmatrix}, \tag{4}$$

calculated in a following sequence:

$$\underset{ordered}{exp}\left[\int_{t_1}^{t_2} \mathbf{D}(t)\right]dt = \lim_{N\to\infty} \underset{ordered}{\prod_{n=1}^{N}} M_n \equiv M_N \cdots M_2 M_1; \Delta t = \frac{t_2 - t_1}{N}; t_n^* \in \{t_1 + (n-1)\Delta t, t_1 + n\Delta t\}$$

$$M_n = \exp[\mathbf{D}(t_n^*)\Delta t] = \begin{bmatrix} \cos\Delta\varphi_n & \sin\Delta\varphi_n / \omega_p(t_n^*) \\ -\omega_p(t_n^*)\sin\Delta\varphi_n & \cos\Delta\varphi_n \end{bmatrix}; \Delta\varphi_n = \omega_p(t_n^*)\Delta t \tag{5}$$

with $\det \mathbf{M}(t_1|t_2) = \exp\int_{t_1}^{t_2} Trace[\mathbf{D}(t)]dt = 1$. In spite of the "focusing" appearance of equation (3), the resulting matrix can correspond to either stable or unstable oscillations.

---

[2] In this paper we are using SGS units. While using $e$ and $m$ is typical for describing electron or proton beams, for charged ion beams one should replace them with $q=Ze$ and $M=Am$ using the charge state $Z$ and atomic number $A$ of the ion.

Periodic modulation of oscillator frequency could result in exponential growth of oscillation amplitude - the phenomena well known as the parametric resonance [79]. Both the width and growth rate of this parametric instability can be enormous when the span of the frequency modulation is large. The well-known example of such instability is a system of periodic focusing lenses with focal lengths ($F$) less than a quarter of distance between them ($L$). The instability is such system has growth rate per cell of [78]:

$$\lambda = -\left(\frac{L}{2F}-1\right)\left(1+\sqrt{1-\left(\frac{L}{2F}-1\right)^{-2}}\right),$$

which be arbitrary large for $L > 4F$. Such exponential instabilities are also well known in accelerators, where a solution of $s$-dependent Hill's equation [75-78]:

$$\frac{d^2x}{ds^2} + K_1(s)x = 0$$

can be unstable even for non-negative $K_1(s) \geq 0$.

These examples prove that modulation of the beam parameters can result in an instability of plasma oscillation and growth of the density modulation in the beam, e.g., in an instability. The beam density modulation can be the result from the changes in the transverse distribution (e.g., from transverse focusing or defocusing forces), from the bunch [3] compression or decompression, or from changes in the beam energy. It is obvious that changes in transverse and longitudinal sizes of the bunch results in the change of its volume and in the corresponding changes in the particles density.

In contrast, changes in the beam energy do not affect particle's density in the laboratory frame but affect mobility of the particles. Because of the Lorentz transformation, in the instantaneous co-moving frame the beam acceleration leads to bunch elongation, increase of distance between particles, and corresponding reduction in their density. Naturally, the bunch deceleration has the opposite effect.

Using the comoving beam frame is very attractive because the traditional set of Vlasov-Maxwell equations is reduced to non-relativistic case of Vlasov-Poisson equations, which is easier to solve. But unfortunately, in the case of beam change its energy (or bending its trajectory the co-moving is no longer inertial - with well-known unpleasant consequences.

Hence, in this paper we will use laboratory frame for our general theory and will use the co-moving frame only for specific cases.

### III.  3D theory of Plasma-Cascade Instability

Let's consider bunch of particles whose motion is described using a standard accelerator coordinate system:

---

[3] Here we use traditional definition for bunch as an assemble of particles limited in all spatial dimensions. In contrast, a beam can be either a sequence of bunches or a continuous flow of particles.

$$\vec{r} = \vec{r}_o(s) + q_1 \cdot \vec{n}(s) + q_2 \cdot \vec{b}(s); \quad \vec{\tau} = \frac{d\vec{r}_o}{ds}; \quad \vec{n} = -\frac{d\vec{r}_o}{ds} \bigg/ \left|\frac{d\vec{r}_o}{ds}\right|; \quad \vec{b} = [\vec{n} \times \vec{\tau}];$$

$$\frac{d\vec{\tau}}{ds} = -K_o(s) \cdot \vec{n}; \quad \frac{d\vec{n}}{ds} = K_o(s) \cdot \vec{\tau} - \kappa_o(s) \cdot \vec{b}; \quad \frac{d\vec{b}}{ds} = \kappa_o(s) \cdot \vec{n};$$

(6)

where $\vec{r}_o(s)$ is trajectory of the reference particle, $K_o(s) = 1/\rho(s)$ is the curvature of the trajectory, and $\kappa_o(s)$ is its torsion. In this case the length along (azimuth) the reference trajectory $s = \int |d\vec{r}_o|$ serves as independent variable [76,80-82] with accelerator Hamiltonian of:

$$H^* = -(1+K_o q_1)\sqrt{\frac{(H-e\varphi)^2}{c^2} - m^2 c^2 - \left(p_1 - \frac{e}{c}A_1\right)^2 - \left(p_2 - \frac{e}{c}A_2\right)^2}$$

$$-\frac{e}{c}A_3 + \kappa_o q_1\left(p_2 - \frac{e}{c}A_2\right) - \kappa_o q_2\left(p_1 - \frac{e}{c}A_1\right);$$

(7)

$$A_3 = (1+K_o q_1)A_s + \kappa(q_2 A_1 - q_1 A_2);$$

and the set of three Canonical pairs of variables [4]

$$\{q_1, p_1\}, \{q_2, p_2\}, \{-ct, p_{ct}\};$$

$$p_1 = p_{m1} - \frac{e}{c}A_1; \quad p_2 = p_{m2} - \frac{e}{c}A_2; \quad p_{ct} = \frac{\mathbf{E} + e\varphi}{c};$$

(8)

where $t$ is arrival time of particle to the azimuth $s$, $\vec{p}_m = \hat{n} \cdot p_{1m} + \vec{b} \cdot p_{2m} + \vec{\tau} \cdot p_{sm}$ and $\mathbf{E}$ are particle's mechanical momentum and energy, correspondingly, and $\{\varphi, \vec{A}\}$ is the 4-potential of EM field. Introduction for the "paraxial" canonical pair:

$$\{q_3, p_3\}; q_3 = c(t_o(s) - t); p_3 = \frac{\mathbf{E} - \mathbf{E}_o(s)}{c} + e\frac{\varphi(\vec{r},t) - \varphi(\vec{r}_o(s),t)}{c}$$

(9)

reduces the Hamiltonian (7) by

$$\frac{c}{v_o(s)} p_3 - q_3 \frac{d}{ds}\left(\mathbf{E}_o(s) + e\frac{\varphi(\vec{r}_o(s),t)}{c}\right),$$

(10)

where we used obvious $\frac{dt_o(s)}{ds} \equiv \frac{1}{v_o(s)}$ and

$$\mathbf{E}_o(s) \equiv \gamma_o(s) mc^2 = c\sqrt{p_o^2(s) + m^2 c^2}; \quad \beta_o(s) \equiv \frac{v_o(s)}{c} = \sqrt{1 - \gamma_o^{-2}(s)}.$$

(11)

---

[4] It is important to note that this is curvilinear coordinate system with neither unit or diagonal metric tensor [76,80-82].

For compactness of our expressions, we will define set of coordinates, $q$, and corresponding Canonical momenta, $p$, as well as the phases space vector $\xi$:

$$q = \{q_1, q_2, q_3\}; \; p = \{p_1, p_2, p_3\}; \; \xi^T = \{q, p\}; \qquad (12)$$
$$\vec{q} = \hat{e}_1 q_1 + \hat{e}_2 q_2 + \hat{e}_3 q_3; \; \vec{p} = \hat{e}_1 p_1 + \hat{e}_2 p_2 + \hat{e}_3 p_3; \; \hat{e}_1 = \vec{n}; \hat{e}_2 = \vec{b}; \hat{e}_3 = \vec{\tau}$$

for 3D case. Number of components are proportionally reduced for 2D and 1D cases. Equations of motion can be rewritten in a symplectic form as [82] [5]

$$\xi^T = [q, p]; \frac{d\xi_i}{ds} = S_{ik} \frac{\partial H}{\partial \xi_k} \Leftrightarrow \frac{d\xi}{ds} = \mathbf{S} \frac{\partial H}{\partial \xi};$$

$$\mathbf{S} \equiv [S_{ik}] = \begin{bmatrix} \mathbf{0} & \mathbf{I}_{3\times 3} \\ -\mathbf{I}_{3\times 3} & \mathbf{0} \end{bmatrix}; \mathbf{I}_{3\times 3} = \begin{bmatrix} 1 & 0 & 0 \\ 0 & 1 & 0 \\ 0 & 0 & 1 \end{bmatrix}; \; \mathbf{S}^2 = -\mathbf{I}_{6\times 6}; \qquad (13)$$

where $\mathbf{0}$ is 3x3 zero matrix (see Appendix A for further discussion) and index $^T$ indicates transposition of matrices, including transferring a column in a row and vice versa.

Motion of particles is determined by initial conditions[6]

$$Q \equiv q(s=0); \; P \equiv p(s=0) \Leftrightarrow \vec{Q} \equiv \vec{q}(s=0); \; \vec{P} \equiv \vec{p}(s=0).$$

Solved equations of motion

$$q = q(Q, P, s); \; p = p(Q, P, t)s; \; \xi^T(X, s) = [q, p]; X^T = [Q, P] \qquad (14)$$

represent the Canonical transformation from $\{Q,P\}$ to $\{q,p\}$ and vice versa. It means that inverse transformation

$$Q = Q(q, p, s); \; P = P(q, p, s); \; X = \{Q, P\} = X(\xi, t) \qquad (15)$$

exists and also represents the Canonical transformation from $\{q,p\}$ to $\{Q,P\}$. The transformation (15) results in trivial Hamiltonian system with set of canonical variables $\{Q, P\}$ and a zero Hamiltonian:

$$\breve{H}(Q, P, s) = 0. \,^7 \qquad (16)$$

---

[5] Further in the paper we will use Einstein's convention of summation by repeated indices, e.g., $a_i b_i \equiv \sum_i a_i b_i; \; a_i b_{ki} c_{nk} \equiv \sum_i \sum_k a_i b_{ki} c_{nk}$.

[6] We will interchangeably use both vector and functional appearances of the coordinates and momenta for compactness of formulae.

[7] Generally speaking, this transformation can leave $\breve{H}(Q, P, s) = f(s).$ , which easily can be removed by a trivial Canonical transformation $F = Q_i P_i - \int\limits^s f(z) dz$.

Since we are considering instability in charge particle beams, we assume that solution for an unperturbed distribution function $F_o$ is known and it satisfies the self-consistent Vlasov equation [8] [83]:

$$\frac{\partial F_o(\xi,s)}{\partial s} + S_{ik}\frac{\partial F_o(\xi,s)}{\partial \xi_i}\frac{\partial H_o(\xi,s)}{\partial \xi_k} = 0;$$

$$\breve{F}_o(X) \equiv F_o(X, s=0) \Rightarrow F_o(\xi,s) = \breve{F}_o(X(\xi,s)).$$

(17)

Let's now consider an infinitesimally small perturbation of the distribution function, $\tilde{f}$,

$$F(\xi,s) = F_o(\xi,s) + \tilde{f}(\xi,s); \ |\tilde{f}(\xi,s)| \ll |F_o(\xi,s)|,$$

(18)

e.g., $\tilde{f}(\xi,s) = O(\varepsilon)|F_o(\xi,s)|; \ \varepsilon \ll 1$, and the corresponding weak perturbation in the Hamiltonian:

$$H(\xi,s) = H_o(\xi,s) + \tilde{h}(\xi,s); \ \tilde{h}(\xi,s) = O(\varepsilon)|H_o(\xi,s)|.$$

(19)

Applying Canonical transformation (15) we reduce the Hamiltonian (19) to the perturbation term

$$\breve{H}(X,s) = \breve{h}(X,s) \equiv \tilde{h}(\xi(X,s),s),$$

(20)

with Vlasov equations for the corresponding variation of the initial distribution function $\breve{f}$:

$$\breve{F}(X,s) = \breve{F}_o(X) + \breve{f}(X,s); \ \tilde{f}(\xi,s) \equiv \breve{f}(X(\xi,s),s);$$

$$\frac{\partial \breve{f}}{\partial s} + S_{ik}\frac{\partial F_o}{\partial X_i}\frac{\partial \breve{h}}{\partial X_k} + S_{ik}\frac{\partial \breve{f}}{\partial X_i}\frac{\partial \breve{h}}{\partial X_k} = 0.$$

(21)

This method, called "the variation on initial values" in analytical mechanics [84] or as "the method of trajectories" in plasma physics [74], is well known. By assuming that solutions for self-consistent trajectories in eqs. (13-15) are know, it allows us to remove (at least formally) dynamic terms and to reduce the Vlasov equations to one comprising only of the perturbation terms. Next standard step is the linearization of the Vlasov equation by recognizing that third term in eq. (21) has order of $O(\varepsilon^2)$:

$$\frac{\partial \breve{f}}{\partial s} + S_{ik}\frac{\partial F_o}{\partial X_i}\frac{\partial \breve{h}}{\partial X_k} = S_{ik}\frac{\partial \breve{f}}{\partial X_i}\frac{\partial \breve{h}}{\partial X_k} = O(\varepsilon^2) \to 0;$$

$$\frac{\partial \breve{f}}{\partial s} + \frac{\partial F_o}{\partial Q_i}\frac{\partial \breve{h}}{\partial P_i} - \frac{\partial F_o}{\partial P_i}\frac{\partial \breve{h}}{\partial Q_i} = 0.$$

(22)

---

[8] Self-consistent distribution function, which we use as the known background, would include all macroscopic collective effects such as space-charge and wake-fields induced by the bunch. Generally speaking, the self-consistent Hamiltonian would have functional dependence on the initial beam distribution $\breve{F}_o(X)$, e.g., $H = H_o(q,p,s,\breve{F}_o(X))$. This fact does not change validity and applicability of the Vlasov equation (17).

A number of further assumptions are needed to derive either analytically solvable equations or those reducible to a directly solvable set of ordinary differential or integral equations [9]. It is well known that a generic 3D evolution of a finite size charged beam is analytically intractable. Rare exceptions, such as non-physical but self-consistent Kapchinsky-Vladimirsky (KV) distribution [85], only attest to the case.

One typical simplification used in the theory of beam instabilities is an assumption of an infinite homogenous plasma. While this approach is not applicable for all collective effects in a beam with finite sizes, it has limited applicability for analyzing evolution of perturbation with periods significantly smaller than typical scales of density uniformity.

It is intuitively understandable that scales of the beam uniformity $a_i$

$$\left|\frac{\partial F_o}{\partial q_i}\right| \propto \frac{F_o}{a_i}$$

define the scale of the perturbations when the infinite homogenous plasma methods can be used as a good approximation. Detailed studies of this approximation are presented in Appendix B and can be summarized as following: for 3D Fourier components, the *k*-vector must satisfy all of the following conditions:

$$\vec{k} = \vec{k}_\perp + \hat{e}_3 k_3; \; a_{1,2} \cdot \sqrt{\vec{k}_\perp^2 + \frac{k_3^2}{\gamma^2}} \gg 1; \; a_3 \cdot \sqrt{\gamma^2 \vec{k}_\perp^2 + k_3^2} \gg 1; \; \gamma = \frac{E}{mc^2}, \quad (23)$$

which include scaling of the longitudinal component of the wavevector ($k_3$) and the bunch length ($a_3$) with the relativistic factor $\gamma$. Since we considering a generic beam transport system, which can include beam's acceleration, compression or decompression, focusing and bending of its trajectory, we shall also assume that changes in the beam and the accelerator parameters at the scale of the density modulation are negligible:

$$\left|\vec{\nabla} g\right| \ll \left|\vec{k}\right| |g| \quad (24)$$

where *g* is an any generic parameter of the accelerator, including but not limited to the beam's energy, velocity, sizes, the accelerator EM fields, the curvature and the torsion of the reference beam trajectory.

Next step in evaluating the instability requires linearization of the symplectic map $\xi = \mathbf{M} : X$ using *6x6* symplectic transport matrix [76,80-82] [10]:

$$\xi = \mathbf{M}(s) X; \; X = \mathbf{M}^{-1}(s)\xi; \; \mathbf{M}(0) = \mathbf{I}_{6\times 6};$$
$$\mathbf{M}^T \mathbf{S} \mathbf{M} = \mathbf{M} \mathbf{S} \mathbf{M}^T = \mathbf{S} \; \Rightarrow \; \det \mathbf{M} = 1; \; \mathbf{M}^{-1} = -\mathbf{S} \mathbf{M}^T \mathbf{S}, \quad (25)$$

which should be evaluated self-consistently, including macroscopic collective effects. It is convenient to identify four *3x3* block-matrices in the transport matrix:

---

[9] Typically, the combination of Vlasov and Maxwell equations is not directly solvable because it contains partial derivatives.
[10] Nonlineary of the map would result in nonlinear, position depentend transformation of the *k*-vector canceling advantages offered by Fourier transformation.

$$\left[\begin{array}{c} q \\ p \end{array}\right] = \mathbf{M}(s)\left[\begin{array}{c} Q \\ P \end{array}\right]; \left[\begin{array}{c} Q \\ P \end{array}\right] = \mathbf{M}^{-1}(s)\left[\begin{array}{c} q \\ p \end{array}\right];$$

$$\mathbf{M} = \left[\begin{array}{cc} \mathbf{A} & \mathbf{B} \\ \mathbf{C} & \mathbf{D} \end{array}\right]; \mathbf{M}^{-1} = -\mathbf{S}\mathbf{M}^T\mathbf{S} = \left[\begin{array}{cc} \mathbf{D}^T & -\mathbf{B}^T \\ -\mathbf{C}^T & \mathbf{A}^T \end{array}\right];$$
(26)

which provide explicit connections between the local and initial coordinates and momenta:

$$q = \mathbf{A}Q + \mathbf{B}P; p = \mathbf{C}Q + \mathbf{D}P;$$
$$Q = \mathbf{D}^T q - \mathbf{B}^T p; P = -\mathbf{C}^T q + \mathbf{A}^T p.$$
(27)

It worth noticing that in this notation three degrees of motion are decoupled when all four 3x3 matrices, **A, B, C** and **D** are diagonal (see Appendix A for more details).

Matrix **A** plays especial role for this instability since is determinant represent the degree of the three-dimensional bunch compression:

$$\frac{d^2 \tilde{\rho}(\vec{q})}{ds^2} + k_p^2 \tilde{\rho}(\vec{q}) = 0; \tilde{\rho}(\vec{q}) = \tilde{\rho}_o(\vec{q})e^{i(k_p s - \omega_{pl} t)}; \omega_{pl} = c\beta_o k_p.$$
(28)

where used one of eq. (27) to connect local beam densities (at azimuth $s$) with their initial values at $s=0$:

$$P = -\mathbf{C}^T q + \mathbf{A}^T p \Rightarrow p = \left(\mathbf{A}^T\right)^{-1}\left(P + \mathbf{C}^T q\right);$$

$$dp^3\bigg|_{q=const} = \frac{1}{\det \mathbf{A}} d\left(P + \mathbf{C}^T q\right)^3 \Rightarrow \frac{1}{\det \mathbf{A}} \int_{-\infty}^{\infty}\int_{-\infty}^{\infty}\int_{-\infty}^{\infty} d\left(P + \mathbf{C}^T q\right)^3 = \frac{1}{\det \mathbf{A}};$$

$$\mathbf{A}Q + \mathbf{B}P = q = const \Rightarrow Q = \mathbf{A}^{-1}(q - \mathbf{B}P);$$

$$F\left(q,\left(\mathbf{A}^T\right)^{-1}\left(P + \mathbf{C}^T q\right)\right) = F_o\left(\mathbf{A}^{-1}(q - \mathbf{B}P), P\right).$$
(29)

One of important consequences of using the assumption of the infinite homogeneous plasma results in requirement of $\det \mathbf{A} > 0$. Otherwise, because of the infinite size of the plasma, beam density would become infinitely large, e.g. unphysical. This is extremely simple to show for the

1D case: $\mathbf{M} = \left[\begin{array}{cc} m_{11} & m_{12} \\ m_{21} & m_{22} \end{array}\right]; \mathbf{M}^{-1} = -\mathbf{S}\mathbf{M}^T\mathbf{S} = \left[\begin{array}{cc} m_{22} & -m_{12} \\ -m_{21} & m_{11} \end{array}\right],$

when $m_{11}$ plays the role of the $\det \mathbf{A}$ and the change in the line density can be easily expressed as

$$\rho(q,s) = \int_{-\infty}^{\infty} F_o(-m_{21}q + m_{11}p) dp = \frac{1}{m_{11}} \int_{-\infty}^{\infty} F_o(p) dp = \frac{n_o}{m_{11}}.$$

It is easy to show that this is no longer a problem for a beam with finite sizes and finite emittances.

As shown in Appendices C, D and E, density perturbation will generate additional potentials of the EM field resulting in perturbation of the accelerator Hamiltonian (see equations (E5)):

$$\tilde{h} = \frac{4\pi e^2}{c} \int \frac{\tilde{\rho}_{\vec{k}} e^{i\vec{k}\vec{q}} dk^3}{\gamma_o^2 \beta_o^2 \vec{k}_\perp^2 + k_3^2}; \quad \delta \frac{d\vec{P}}{dt} = -\frac{\partial \tilde{h}}{\partial \vec{q}} = -\frac{4\pi e^2}{c} \int \frac{i\vec{k} \cdot \tilde{\rho}_{\vec{k}} e^{i\vec{k}\vec{q}} dk^3}{\gamma_o^2 \beta_o^2 \vec{k}_\perp^2 + k_3^2}. \quad (30)$$

We can easily connect the $\tilde{\rho}_{\vec{k}}$ at location $s$ with Fourier harmonic of $\breve{f}$. Taking into account conservation of the phases-space volume $dq^3 dp^3 = \det \mathbf{M} \cdot dQ^3 dP^3 = dQ^3 dP^3 \; dQ^3 dP^3 \equiv dq^3 dp^3$ and conservation of the phase space density $\breve{f}(X,s) \equiv \tilde{f}(\xi(X,s),s)$ we get:

$$\tilde{\rho}_{\vec{k}} \equiv \tilde{\rho}(s,\vec{k}(s)) = \frac{1}{(2\pi)^3} \int\int^3 \breve{f}(Q,P,s) e^{-i\vec{k}(s)\vec{q}(X,t)} dQ^3 dP = \frac{1}{(2\pi)^3} \int\int \breve{f} \cdot e^{-i\left(\vec{k}_o \cdot \vec{Q} + \vec{k}_o \cdot \overrightarrow{\mathbf{A}^{-1}\mathbf{B}} \cdot \vec{P}\right)} dQ^3 dP^3, \quad (31)$$

where we used $\vec{q} = \vec{\mathbf{A}} \cdot \vec{Q} + \vec{\mathbf{B}} \cdot \vec{P}$ as equivalent of $q = \mathbf{A}Q + \mathbf{B}P$ in eq. (27), a compact notation for convolution of two vectors and a matrix:

$$\vec{x} \cdot \vec{\mathbf{A}} \cdot \vec{y} = \sum_{i=1}^{3}\sum_{j=1}^{3} A_{ij} x_i y_j$$

and explicit indication that the k-vector $\vec{k}(s)$ is function of $s$:[11]

$$\vec{k}(s) = \vec{k}_o \cdot \mathbf{A}^{-1}(s); \; \vec{k}_o = \vec{k}(s=0); \; \vec{k}(s) \cdot \vec{q} = \vec{k}_o \cdot \vec{Q} + \vec{k}_o \cdot \overrightarrow{\mathbf{A}^{-1}\mathbf{B}} \cdot \vec{P}. \quad (32)$$

It means that matrix $\mathbf{A}$, the spatial components of the transport matrix, also defines evolution of the $k$-vector with initial value of $\vec{k}_o$:

$$k^T(s) = [k_1(s), k_2(s), k_3(s)]; \; k_o = \mathbf{A}^T k(s) \Leftrightarrow k(s) = (\mathbf{A}^T)^{-1} k_o. \quad (33)$$

We can assume, without loss of generality, that initial distribution is an arbitrary integrable function of momenta [12]:

$$F_o = n_o F_o(P); \int_{-\infty}^{\infty} F_o(P) dP^3 = 1; \; n_o = \frac{j_o}{ec}, \quad (34)$$

where $j_o$ is the initial beam current density. It is important to note that in contrast with velocity-dependent spatial density of the beam, $n_l = j_o / ev_o$, the $n_o = \beta_o n_r$ has a well-defined finite value. Now we can rewrite Vlasov equation (22) using relations from eq. (29):

---

[11] For compactness, in places where it cannot cause confusion, we omit explicit indication $s$-dependence, for example using $\overleftrightarrow{\mathbf{A}^{-1}\mathbf{B}}$ instead of $\overleftrightarrow{\mathbf{A}(s)^{-1}\mathbf{B}(s)}$ in this equation.

[12] For plasma to remain uniform the distribution must have form of $f(p+\mathbf{M}q)$. Initial linear correlations between $p$ and $q$ can be incorporated into the transport matrix (26).

$$P = -\mathbf{C}^T q + \mathbf{A}^T p; \; dP_i = A_{ji} dp_j - C_{ji} dq_j;$$

$$\frac{\partial \breve{f}}{\partial s} = -n_o \frac{\partial F_o}{\partial P_i} \frac{\partial P_i}{\partial p_j} \delta\left(\frac{dp_j}{ds}\right) - n_o \frac{\partial F_o}{\partial P_i} \frac{\partial P_i}{\partial q_j} \delta\left(\frac{dq_j}{ds}\right) = n_o \frac{\partial F_o}{\partial P_i} A_{ji} \frac{\partial \tilde{h}}{\partial q_j};$$

and taking into account that

$$\delta\left(\frac{dq_j}{ds}\right) = \frac{\partial \tilde{h}(\xi,s)}{\partial p_j} = 0.$$

Using (30) we arrive to self-consistent Vlasov equations:

$$\frac{\partial \breve{f}}{\partial s} = n_o \frac{\partial F_o}{\partial P_i} A_{ji}(s) \mathbf{F}_j(q,s); \; \vec{\mathbf{F}}(q,s) = \frac{\partial \tilde{h}}{\partial q_j} = \frac{4\pi e^2}{c} \int \frac{i\vec{k} \cdot \tilde{\rho}_{\vec{k}} e^{i\vec{k}\vec{q}} dk^3}{\gamma_o^2 \beta_o^2 \vec{k}_\perp^2 + k_3^2}, \tag{35}$$

which is suitable for the Fourier transform $f_{\vec{k}_o}(P,s) = \frac{1}{2\pi} \int\limits_{-\infty}^{\infty} f(Q,P,s) e^{-i\vec{k}_o \vec{Q}} dQ^3$:

$$\frac{\partial \breve{f}_{\vec{k}_o}}{\partial s} = n_o \frac{\partial F_o}{\partial P_i} A_{ji}(t) \int dQ^3 e^{-i\vec{k}_o \vec{Q}} \mathbf{F}_j(q,t). \tag{36}$$

The later has to be evaluated at $\vec{P} = const$ using established relations between $k$-vectors (33):

$$\mathbf{F}_{\vec{k}'} = \int e^{-i\vec{k}_o \vec{Q}} \mathbf{F}(q,s) dQ^3 \bigg|_{P=const} = \frac{4\pi e^2}{c} \int \frac{i\vec{k} \tilde{\rho}_{\vec{k}} dk^3}{\gamma_o^2 \beta_o^2 \vec{k}_\perp^2 + k_3^2} \frac{1}{(2\pi)^3} \int e^{i\vec{k}\vec{q}} e^{-i\vec{k}_o \vec{Q}} dQ^3;$$

$$\frac{1}{(2\pi)^3} \int e^{i\vec{k}\vec{q}} e^{-i\vec{k}_o \vec{Q}} dQ^3 = \frac{e^{i\vec{k}\cdot\overset{\leftrightarrow}{\mathbf{B}}\cdot\vec{P}}}{(2\pi)^3} \int e^{i(\vec{k}\cdot\vec{\mathbf{A}}-\vec{k}_o)\vec{Q}} dQ^3 = e^{i\vec{k}\cdot\vec{\mathbf{B}}\cdot\vec{P}} \delta(\vec{k}\cdot\vec{\mathbf{A}} - \vec{k}_o) = \frac{e^{i\vec{k}_o \cdot \overline{\mathbf{A}^{-1}\mathbf{B}} \cdot \vec{P}}}{\det \mathbf{A}} \delta(\vec{k} - \vec{k}_o \vec{\mathbf{A}}^{-1}),$$

resulting in

$$\frac{\partial \breve{f}_{\vec{k}_o}(P,s)}{\partial s} = \frac{4\pi n_o e^2}{c} \cdot \frac{\tilde{\rho}(s,\vec{k}(s))}{\gamma_o(s)^2 \beta_o(s)^2 \vec{k}_\perp(s)^2 + k_3(s)^2} \frac{e^{i\vec{k}_o \cdot \overline{\mathbf{A}^{-1}(s)\mathbf{B}(s)} \cdot \vec{P}}}{\det \mathbf{A}(s)} \left(ik_{oi} \frac{\partial F_o}{\partial P_i}\right), \tag{37}$$

where we took inti account that $k_j(s) A_{ji}(s) \frac{\partial F_o}{\partial P_i} = k_o \frac{\partial F_o}{\partial P_i}$. This equation can be integrated:

$$\breve{f}_{\vec{k}_o}(P,s) = \breve{f}_{\vec{k}_o}(P,0) + \frac{4\pi i n_o e^2}{c} \left(k_{oi} \frac{\partial F_o}{\partial P_i}\right) \int\limits_0^s \frac{e^{i\vec{k}_o \cdot \overline{\mathbf{A}^{-1}(\zeta)\mathbf{B}(\zeta)} \cdot \vec{P}}}{\det \mathbf{A}(\zeta)} \frac{\tilde{\rho}_{\vec{k}}(\zeta) d\zeta}{\gamma_o(\zeta)^2 \beta_o(\zeta)^2 \vec{k}_\perp^2(\zeta) + k_3^2(\zeta)}. \tag{38}$$

Rewriting (31) as

$$\tilde{\rho}(s,\vec{k}(s)) = \frac{1}{(2\pi)^3} \int \breve{f}_{\vec{k}_o}(P,s) \cdot e^{-i\vec{k}_o \cdot \overline{\mathbf{A}(s)^{-1}\mathbf{B}(s)} \cdot \vec{P}} dP^3, \tag{39}$$

turns eq. (38) into a directly solvable integral equation:

$$\tilde{\rho}\left(s,\vec{k}(s)\right)=\tilde{\rho}_{\vec{k}0}(s)+\frac{4\pi ie^2 n_o}{c}\int_o^s \frac{\tilde{\rho}\left(\zeta,\vec{k}(\zeta)\right)d\zeta}{\det \mathbf{A}(\zeta)}\int \frac{e^{i\vec{k}_o\cdot(\vec{U}(\zeta)-\vec{U}(s))\cdot\vec{P}}}{\gamma_o(\zeta)^2 \beta_o(\zeta)^2 \vec{k}_\perp^2(\zeta)+k_3^2(\zeta)} k_{oi}\frac{\partial F_o}{\partial P_i}dP^3; \quad (40)$$

$$\mathbf{U}(s)=\mathbf{A}^{-1}(s)\mathbf{B}(s);\ \tilde{\rho}_{\vec{k}0}(s)=\int e^{-i\vec{k}_o\cdot\vec{U}(s)\cdot\vec{P}}\tilde{f}_{\vec{k}_o}(P,0)dP^3.$$

While this equation already can be used for evaluation of the instability, it can be further simplified by eliminating convolution $\sum_{i=1}^{3}\frac{\partial F_o}{\partial P_i}k_i'$. Using integration by parts

$$\int \frac{\partial F_o}{\partial P_i}\phi dP_i = F_o \phi \Big|_{P_i=-\infty}^{P_i=\infty} - \int F_o \frac{\partial \phi}{\partial P_i}dP_i$$

and $F_o(P_i=\pm\infty)=0$ we get:

$$\sum_{i=1}^{3}k_{oi}\frac{\partial}{\partial P_i}e^{i\vec{k}_o\cdot(\vec{U}(\zeta)-\vec{U}(s))\cdot\vec{P}} = -ik_o^2\left(u(s)-u(\zeta)\right);$$

$$k_o=\left|\vec{k}_o\right|;\ \vec{k}_o=k_o\vec{v}_o;\ u(s)=\vec{v}_o\cdot\vec{\vec{U}}(\zeta)\cdot\vec{v}_o \equiv \sum_{i,j}\mathbf{U}_{ij}(\zeta)\cdot v_{0i}v_{oj}, \quad (41)$$

where we introduced unit vector $\vec{v}_o$ in the direction of the initial *k*-vector. If is convenient to extent is definition of this dimensionless vector as

$$\vec{k}(s) = k_o\vec{v}(s);\ \vec{v}(s) = \vec{k}_o\vec{\vec{A}}^{-1}(s). \quad (42)$$

We show in eq. (A12) of Appendix A that $\mathbf{AB}^T = \mathbf{BA}^T$, which also means that $\mathbf{U}=\mathbf{A}^{-1}\mathbf{B}$ is also a symmetric matrix.

Combining eqs. (40) and (41) This brings us to the final form of integral equation for this instability:

$$\tilde{\rho}\left(s,\vec{k}(s)\right)=-\int_o^s \tilde{\rho}\left(\zeta,\vec{k}(\zeta)\right)\cdot K(\zeta)\left(u(s)-u(\zeta)\right)L_d(s,\zeta)d\zeta + \tilde{\rho}_{o\vec{k}}(s);$$

$$K(\zeta)=\frac{4\pi n_o e^2}{c\det \mathbf{A}(\zeta)v(\zeta)};\ L_d(k_o,s,\zeta)=\int e^{i(\vec{\kappa}(\zeta)-\vec{\kappa}(s))\cdot\vec{P}}F_o(P)dP^3; \quad (43)$$

$$\vec{\kappa}(\zeta)=\vec{k}_o\cdot\vec{\vec{U}}(\zeta)\equiv k_o\vec{v}_o\vec{\vec{U}}(\zeta);\ v(s)=\gamma_o(s)^2\beta_o(s)^2\vec{v}_\perp(s)^2+v_3(s)^2;$$

which can be solved numerically for any accelerator.

It is important to note that in the kernel of the integral equation (43) there in only one term, the Landau damping [87], $L_d$, depend on the absolute value of the *k*-vector. The rest of the terms, $K$ and $u$, are defined by the geometry (e.g. direction of the initial *k*-vector), the components of accelerator transport matrix in form of matrix $\mathbf{U}=\mathbf{A}^{-1}\mathbf{B}$ and s-dependent denominator $\det \mathbf{A}(\zeta)\upsilon(\zeta)$. The most non-trivial construction is actually $\upsilon(\zeta)$, which is the result of the asymmetry introduced by Lorentz transformation of the 4-potential:

$$\upsilon(s) = \vec{v} \cdot \vec{\mathbf{G}}(s) \cdot \vec{v}; \vec{\mathbf{G}} = \mathbf{A}^{-1} \begin{bmatrix} \gamma_o^2 \beta_o^2 & 0 & 0 \\ 0 & \gamma_o^2 \beta_o^2 & 0 \\ 0 & 0 & 1 \end{bmatrix} (\mathbf{A}^{-1})^T. \quad (44)$$

Furthermore, the convolution $u(s) = \vec{v}_o \cdot \vec{\mathbf{U}}(\zeta) \cdot \vec{v}_o$ (41) has important non-trivial properties that it is nonnegative monotonic function with positive derivative (see eq. (A16) in Appendix A):

$$u(s) \geq 0; \; u'(s) > 0. \quad (45)$$

Generally speaking, for a beam with arbitrary momentum spread the equation (43) cannot be either evaluated analytically or reduced in complexity. But physical nature of various terms can be identified by considering specific cases. For example, the integral over the momenta is known as Landau damping [87] and can be easily evaluated for Gaussian distribution:

$$F_o(P) = \prod_{i=1}^{3} \frac{1}{\sqrt{2\pi}\sigma_i} \exp\left(-\frac{P_i^2}{2\sigma_i^2}\right) \quad (46)$$

generating exponential term

$$L_d = \int e^{i(\vec{\kappa}(\xi)-\vec{\kappa}(s))\cdot\vec{P}} F_o(P) dP^3 = \prod_{i=1}^{3} \exp\left(-\frac{k_o^2 \sigma_i^2 (\kappa_i(\xi) - \kappa_i(s))^2}{2}\right); \quad (47)$$

corresponding to the decay of the modulation during the interval $(\xi, s)$.

Equation (43) is the most general equation that describes evolution of high-frequency modulation in beams driven by space charge effects. It can be also used to describe one dimensional or 2D instabilities. For example, it is easy to show that conventional longitudinal microwave instability [53-71] can be also described by this equation under simplified assumptions. Specifically, conventional theory of longitudinal microwave instability assumes that in straight sections the longitudinal motion is frozen and energy modulation resulted from accumulated space-charge forces is transferred into density by $R_{56}$ of a magnetic system. Furthermore, space charge is frequently neglected in the bending magnetic system. Hence, eq. (43) is an universal equation for description of instabilities driven by the space charge.

### IV. Specific cases

In this section, we review some specific cases which are of interest for this paper. In some cases, we can simplify eq. (43) or reduce it to second-order ordinary differential equation (ODE).

Let's consider cases when the Landau damping term allows separation of variables $s$ and $\zeta$[13]:

$$L_d(s,\zeta) = \Lambda(\zeta)\Lambda^{-1}(s); \Lambda(s) = e^{-\phi(s)}; \quad (48)$$

and the integral equation (33) becomes:

$$\tilde{q}(s) = -\int_0^s \tilde{q}(\zeta) K(\zeta)(u(s)-u(\zeta))d\zeta + \tilde{q}_o(s); \quad (49)$$

for scaled density modulation $\tilde{q}(s) = e^{\phi(s)}\tilde{\rho}(s,\vec{k}(s))$. Combination of first and second derivatives of eq. (49) transfers it into the second order ODE:

$$\tilde{q}'' - \alpha' \cdot \tilde{q}' + Ku' \cdot \tilde{q} = \tilde{q}_o'' - \alpha' \cdot \tilde{q}_o'; $$
$$\tilde{q}_o(s) = e^{\phi(s)}\tilde{\rho}_{\vec{k}0}(s); \alpha = \ln\frac{u'}{u'_o}; u'_o = u'(0), \quad (50)$$

where we used fact that $u'>0$ and standard notation for derivatives $f' = \frac{df}{ds}; f'' = \frac{d^2f}{ds^2}$. Finally, this equation can be also reduced to inhomogeneous Hill's equation

$$\hat{q}'' + \hat{K}(s)\hat{q} = \varsigma(s); \hat{K}(s) = K(s)u'(s) - \frac{\alpha'(s)^2}{4} + \frac{\alpha''(s)}{2};$$
$$\hat{q} = e^{-\frac{\alpha(s)}{2}}\tilde{q} \equiv \sqrt{\frac{u'_o}{u'(s)}} \cdot e^{\phi(s)} \cdot \tilde{\rho}(s,\vec{k}s) ; \varsigma(s) = e^{-\frac{\alpha(s)}{2}}(\tilde{q}_o''(s) - \tilde{q}_o'(s)\alpha'(s)). \quad (51)$$

It is well known that solution of homogeneous Hill's equation is represented by symplectic matrix (see eqs. (3-5)):

$$\begin{bmatrix} \hat{q}(s) \\ \hat{q}'(s) \end{bmatrix} = \mathbf{R}(s)\begin{bmatrix} \hat{q}(0) \\ \hat{q}'(0) \end{bmatrix}; \mathbf{R} = \begin{bmatrix} r_{11} & r_{12} \\ r_{21} & r_{22} \end{bmatrix}; \mathbf{R}' = \begin{bmatrix} 0 & 1 \\ -\hat{K}(s) & 0 \end{bmatrix}\mathbf{R}; \det \mathbf{R} = 1;, \quad (52)$$

which also defines general solutions of inhomogeneous equation:

$$\hat{q}(s) = r_{11}(s)\hat{q}(0) + r_{12}(s)\hat{q}'(0) + \int_0^s (r_{11}(\zeta)r_{12}(s) - r_{11}(s)r_{12}(\zeta))\varsigma(\zeta)d\zeta. \quad (53)$$

Hence, solution of the homogenous Hill's equation $\hat{q}'' + \hat{K}(s)\hat{q} = 0$ is the key for investigation of this instability, when the separation (48) is possible. Cold beam with momenta distribution

$$F_0(P) = \delta(P_1)\delta(P_2)\delta(P_3)$$

---

[13] Unfortunately, as can be seen from eq. (47), such separation is impossible for Gaussian momenta distribution.

definitely satisfy this requirement with $\phi = 1$. More general and interesting is the case of beam with κ-1 momentum distribution in all directions:

$$F_0(P) = F_{\kappa-1}(P) = \frac{1}{\pi^3} \prod_{i=1}^{3} \frac{\sigma_i}{\sigma_i^2 + P_i^2}, \tag{54}$$

allowing to integrate over the momenta:

$$L_d(\zeta,s) = \int e^{ik_o(\vec{\kappa}(\zeta)-\vec{\kappa}(s))\cdot\vec{P}} F_o(P) dP^3 = e^{-\sum_{i=1}^{3}\sigma_i|\kappa_i(s)-\kappa_i(\zeta)|}. \tag{55}$$

If condition $|\kappa_i(s)| \geq |\kappa_i(\zeta)|$; $s \geq \zeta$ is satisfied for all three components $i = 1, 2, 3$, we can used $|\kappa_i(s) - \kappa_i(\zeta)| = |\kappa_i(s)| - |\kappa_i(\zeta)|$ and separated variables:

$$L_d(s,\zeta) = e^{\phi(\zeta)} e^{-\phi(s)}; \quad \phi(\zeta) = \sum_{i=1}^{3} \sigma_i |\kappa_i(\zeta)|; \tag{56}$$

resulting in a second order ODE (50) for $\tilde{q}_{\vec{k}}(s) = \tilde{\rho}(s, \vec{k}(s)) \cdot \exp(\phi(s))$, and in Hill's equation (51) for $\hat{q}(s) = \tilde{\rho}(s, \vec{k}(s)) \sqrt{u'_o / u'(s)} \cdot \exp(\phi(s))$.

This is a good place to discuss driving term, $\varsigma(s)$, in the right-hand side (r.h.s.) of Hills equation

$$\varsigma(s) = e^{-\frac{\alpha}{2}} (\tilde{q}''_o - \tilde{q}'_o \alpha'); \quad \tilde{q}_o(s) = e^{\phi(s)} \tilde{\rho}_{\vec{k}0}(s) = e^{\phi(s)} \int e^{-i\vec{\kappa}(s)\cdot\vec{P}} \breve{f}_{\vec{k}_o}(P, 0). \tag{57}$$

Generally speaking, for an arbitrary initial perturbation $\breve{f}(P,Q)$, both $\tilde{q}''$ and $\tilde{q}'_o$ are not equal zero and Hill's equation remains inhomogeneous. One case is an exception: when the initial perturbation is fully defined by the density perturbation:

$$\breve{f}(P,Q) = \breve{\rho}_o(Q) \cdot F_{\kappa-1}(P) \rightarrow \int e^{-i\vec{\kappa}(s)\cdot\vec{P}} \breve{f}_{\vec{k}_o}(P,0) = \breve{\rho}_{\vec{k}_o} e^{-\phi(s)};$$
$$\tilde{q}_o(s) = \breve{\rho}_{\vec{k}_o} = \int \breve{\rho}_o(Q) e^{-i\vec{k}_o\cdot Q} = const, \tag{58}$$

all derivatives of $\tilde{q}_o$ are equal zero, and the Hill's equation becomes homogenous:

$$\breve{f}(P,Q) = \breve{\rho}_o(Q) \cdot F_{\kappa-1}(P) \rightarrow \hat{q}'' + \hat{K}(s)\hat{q} = 0. \tag{59}$$

While the conditions $|\kappa_i(s)| \geq |\kappa_i(\zeta)|$; $s \geq \zeta$ are frequently satisfied, they also can be violated in the case of an arbitrary coupling. In fact, it is possible to construct matrix $\mathbf{U}$ that one component of vector $\vec{\kappa} = \vec{k}_o \mathbf{U}$ turns from non-zero value at $\zeta$ to zero at $s > \zeta$. Emittance exchange lattices can serve as an example. If even one of $|\kappa_i(s)| \geq \kappa_i(\zeta)$ conditions is violated, the separation becomes impossible.

As we shown in Appendix A (see eq.(A29)) that in the case of uncoupled motion the matrix $\mathbf{U}$ is diagonal with monotonically growing diagonal terms:

$$\mathbf{U}(s) = [\delta_{ij}] \mu_i(s); \; \mu_i(0) = 0; \mu_i(s) > \mu_i(\zeta) \; \forall \zeta < s \; \delta_{ij}\mu_i$$

which means that

$$|\kappa_i(s)| = |k_{io}|\mu_i(s) \tag{60}$$

are also monotonic functions satisfying conditions $|\kappa_i(s)| \geq |\kappa_i(\zeta)|; \; s \geq \zeta$.

Hence, we proved that in accelerator with decoupled motion one can use second order ODE (50) or Hill's equation (51) for beam with κ-1 momentum (energy) distributions. This also include linear accelerators using solenoids – the equations of motion are decoupled by using torsion $\kappa_o(s) = -\dfrac{eB_s}{2p_o c}$ (see eq. (6) [86]).

For the beam with the constant density and constant energy propagating in a drift space we have

$$\mathbf{A} = \mathbf{D} = \mathbf{I}; \; \mathbf{C} = \mathbf{0}; \; \mathbf{U}(s) = \mathbf{B}(s) = \frac{1}{\gamma_o \beta_o mc} \begin{bmatrix} s & 0 & 0 \\ 0 & s & 0 \\ 0 & 0 & s/\gamma_o^2 \beta_o^2 \end{bmatrix}; \tag{61}$$

$$\vec{k} = const; \; \tilde{\rho}_{0\vec{k}} = const; \; u = \frac{s}{\gamma_o^3 \beta_o^3 mc}\left(\gamma_o^2 \beta_o^2 \vec{k}_\perp^2 + k_3^2\right); \; u'K = \frac{4\pi n_o e^2}{\gamma_o^3 \beta_o^3 mc} = const; \; u'' = 0.$$

For cold plasma oscillations it results in well known $\vec{k}$-independent equation:

$$\frac{d^2 \tilde{\rho}_{\vec{k}}}{ds^2} + k_p^2 \tilde{\rho}_{\vec{k}} = 0; \; k_p^2 = \frac{4\pi n_o e^2}{\gamma_o^3 \beta_o^3 mc}, \tag{62}$$

which, after applying the inverse Fourier transformation, becomes the carbon copy of eq. (2) but in the laboratory frame:

$$\frac{d^2 \tilde{\rho}(\vec{q})}{ds^2} + k_p^2 \tilde{\rho}(\vec{q}) = 0; \; \tilde{\rho}(\vec{q}) = \tilde{\rho}_o(\vec{q}) e^{i(k_p s - \omega_{pl} t)}; \; \omega_{pl} = c\beta_o k_p. \tag{63}$$

For beams with κ-1 momentum distribution (54) propagating in a drift space with the constant density and constant energy, $\vec{k}$-dependence occurs via Landau damping term and $q''_{\vec{k}o}$ as the driving term:

$$\tilde{q}'' + k_p^2(s)\tilde{q} = \tilde{q}_o''; \; \tilde{q} = e^{\phi(s)} \tilde{\rho}_{\vec{k}}; \tilde{q}_o = e^{\phi(s)} \tilde{\rho}_{\vec{k}0};$$

$$\phi(s) = \frac{s}{\gamma_o \beta_o mc} \cdot \left(\sigma_1 |k_1| + \sigma_2 |k_2| + (\gamma_o \beta_o)^{-2} \sigma_3 |k_3|\right), \tag{64}$$

$$\tilde{\rho}_{\vec{k}0} = \int \exp\left(i \frac{s}{\gamma_o \beta_o mc} \cdot \left(k_1 \sigma_1 P_1 + \sigma_2 k_2 P_2 + (\gamma_o \beta_o)^{-2} \sigma_3 k_3 P_3\right)\right) \breve{f}_{\vec{k}_o}(P,0) dP^3.$$

As one important consequence of eq. (64) is that for beam propagating in straight section Landau damping decrement for transverse modulation is boosted by factor $\gamma_o^2\beta_o^2$, which is typically >>1. In other words, for $|k_{1,2}|\sigma_{1,2} \propto |k_3|\sigma_3$, the Landau damping term is significantly larger than for $k_{1,2}=0$. This is one of the reasons why longitudinal PCI is of special interest in the paper.

Let's consider 1D longitudinal instability in a beam propagating along straight trajectory, e.g. when the longitudinal and transverse motion are decoupled:

$$\mathbf{A}(s)=\begin{bmatrix} \mathbf{A}_\perp(s) & 0 \\ 0 & a_\|(s) \end{bmatrix};\ \mathbf{B}(s)=\begin{bmatrix} \mathbf{B}_\perp(s) & 0 \\ 0 & b_\|(s) \end{bmatrix};\ \vec{k}(s)=\hat{e}_3 k(s)=\hat{e}_3 \frac{k_o}{a_\|(s)}. \tag{65}$$

Evolution of this instability can be described either by integral equation:

$$\tilde{\rho}(s,k(s))=-\frac{4\pi n_o e^2}{c}\int_o^s \tilde{\rho}(\zeta,k(\zeta))K_\|(\zeta)(u(s)-u(\zeta))d\zeta \int e^{-ik_o(u(s)-u(\zeta))\cdot P} F_\|(P)dP + \tilde{\rho}_{ok}(s);$$

$$K_\|(\zeta)=\frac{4\pi n_o e^2}{c}\frac{a_\|(\zeta)}{\det \mathbf{A}_\perp(\zeta)};\ u(\zeta)=\frac{b_\|(\zeta)}{a_\|(\zeta)};\ \tilde{\rho}_{ok}(s)=\int e^{-ik_o u(s)P}\breve{f}_{k'\|}(P,0)dP. \tag{66}$$

or for κ-1 longitudinal momentum distribution by differential equation:

$$\tilde{q}''-\xi(s)\cdot\tilde{q}'+k_p^2(s)\tilde{q}=\tilde{q}_o''-\xi(s)\cdot\tilde{q}_o';$$

$$k_p^2(s)=\frac{4\pi}{(\gamma_o\beta_o)^3}\cdot\frac{n_o r_c}{a_\| \det \mathbf{A}_\perp};\ \xi(s)=\frac{d}{ds}\left(\ln a_\|^2(\gamma_o\beta_o)^3\right);\ \tilde{q}(s)=\tilde{\rho}\left(s,\frac{k_o}{a_\|(s)}\right)e^{\frac{k_o b_\|(s)\sigma_3}{a_\|(s)}}, \tag{67}$$

where $k_p(s)$ is s-depended frequency (is s domain), $k/a_\|(s)$ is scaled wavenumber of the perturbation, and $\xi(s)$ represents an addition term which. depending on tits sign, either damps or amplifies modulation. The corresponding Hill's equation has the same driving term but slightly different s-depended frequency:

$$\hat{q}''+k_p'^2\hat{q}=a_\|(\gamma_o\beta_o)^{3/2}\left(\tilde{q}_o''-\tilde{q}_o'\frac{u''}{u'}\right);\ k_p'^2=k_p^2-\frac{\xi'^2}{4}+\frac{\xi''}{2}. \tag{68}$$

For beam with the constant energy and no-compression ($a_\|=1$) equations (67) and (68) become identical and describe PCI driven only by transverse focusing:

$$a_\|=1;\ \gamma_o\beta_o=const; b_\|=\frac{s}{(\gamma_o\beta_o)^3 mc} \rightarrow \tilde{q}''+k_p^2(s)\tilde{q}=\tilde{q}_o'';$$

$$k_p^2(s)=\frac{4\pi}{(\gamma_o\beta_o)^3}\cdot\frac{n_o r_c}{\det \mathbf{A}_\perp(s)};\ \rho_k(s)=\tilde{q}(s)e^{-\frac{ks}{(\gamma_o\beta_o)^2}\frac{\sigma_{\gamma_o}}{\gamma_o\beta_o}}. \tag{69}$$

## V. Conclusions.

In this paper we developed general self-consistent 3D theory of high-frequency microbunching instability driven by space charge forces. We derived the directly solvable integral equation (43) fully describing any such instability withing well-defined range of assumption summarized by eqs. (23-24). We also derived conditions when this integral equation can be reduced to a second order ODE (51).

This theory is applicable to all accelerators and it accurately describes both the newly discovered PCI and conventional MBI withing the range of assumption summarize in eq. (23). Furthermore, this theory goes beyond traditional model used to describe conventional MBI as alternating pairs "space-charge kick" – "$R_{56}$ drift". Our solution is self-consistent including continuous evolution in all elements of accelerators. As can be seen from eq. (32), bending of the particle trajectory couples longitudinal and transverse modulations resulting in non-trivial evolution in accelerator arcs or chicanes.

The authors would like to thank all of their colleagues from BNL who contributed to the CeC project, with special acknowledgements going to the CeC group and Dr. Thomas Roser for encouragement and unrelenting support of this research. The first author would also like to thank Prof. Pietro Musumeci (UCLA), who mentioned during our discussion that the modulation of the transverse beam size can violate energy conservation in longitudinal plasma oscillations. This notion sparked the initial impulse for us to investigate if the modulation of the transverse beam size could cause a broad-band longitudinal instability.

This research was supported by NSF grant PHY-1415252, by DOE NP office grant DE- FOA-0000632, and by Brookhaven Science Associates, LLC under Contract No. DE-SC0012704 with the U.S. Department of Energy.

.

**Appendix A – System Hamiltonian and equations of motion**

Traditionally in accelerator physics literature the vector X in phase scape is combined from Canonical pairs of coordinates and momenta $X^T = \left[ ...(q_i, P^i)... \right] \equiv \left[ ...(x_{2i-1}, x_{2i})... \right]$. In this paper, following [82], we use equivalent but different structure of the X-vector, which clearly separate coordinates and momenta and simple form of the matrix of symplectic generator, **S**;

$$X^T = [x_1, ..., x_{2n}] = [Q^T, P^T]; Q^T = [q_1, ..., q_n]; P^T = [P^1, ..., P^n];$$

$$\begin{cases} \dfrac{dq_i}{dt} = \dfrac{\partial H}{\partial P^i} \\ \dfrac{dP^i}{dt} = \dfrac{\partial H}{\partial q_i} \end{cases} \Leftrightarrow \dfrac{dX}{dt} = \mathbf{S} \dfrac{dH}{dX} \Leftrightarrow \dfrac{dx_i}{dt} = \mathbf{S}_{ij} \dfrac{dH}{dx_j}; \tag{A1}$$

$$\mathbf{S} \equiv [S_{ik}] = \begin{bmatrix} \mathbf{0} & \mathbf{I} \\ -\mathbf{I} & \mathbf{0} \end{bmatrix}; \mathbf{I}_{n \times n} = \begin{bmatrix} 1 & 0 & 0 \\ 0 & ... & 0 \\ 0 & 0 & 1 \end{bmatrix}.$$

Use of these notations is especially convenient for linear maps in the form of *2nx2n* symplectic transport matrices:

$$X(t_2) \equiv \begin{bmatrix} Q(t_2) \\ P(t_2) \end{bmatrix} = \mathbf{M}(t_1|t_2) X(t_1) \equiv \mathbf{M}(t_1|t_2) \begin{bmatrix} Q(t_1) \\ P(t_1) \end{bmatrix};$$

$$\mathbf{M}^T \mathbf{S} \mathbf{M} = \mathbf{M} \mathbf{S} \mathbf{M}^T = \mathbf{S}; \mathbf{M}^{-1} = -\mathbf{S} \mathbf{M}^T \mathbf{S};$$

$$\mathbf{M} = \begin{bmatrix} \mathbf{A} & \mathbf{B} \\ \mathbf{C} & \mathbf{D} \end{bmatrix}; \mathbf{M}^{-1} = \begin{bmatrix} \mathbf{D}^T & -\mathbf{B}^T \\ -\mathbf{C}^T & \mathbf{A}^T \end{bmatrix}; \tag{A2}$$

$$Q(t_2) = \mathbf{A} Q(t_1) + \mathbf{B} P(t_1); P(t_2) = \mathbf{C} Q(t_1) + \mathbf{D} P(t_1);$$
$$Q(t_1) = \mathbf{D}^T Q(t_2) - \mathbf{B}^T P(t_2); P(t_1) = -\mathbf{C}^T Q(t_2) + \mathbf{A}^T P(t_2);$$

providing explicit connection between coordinates and momenta with their initial values and vice versa. It also provides important properties of the block matrices which can be very useful for the evaluation of complex expression. Specifically, symplecticity of transport matrix requires that four *nxn* matrices $\mathbf{AB}^T, \mathbf{DC}^T, \mathbf{A}^T \mathbf{C}, \mathbf{D}^T \mathbf{B}$ will be symmetric

$$\left(\mathbf{AB}^T\right)^T = \mathbf{AB}^T; \left(\mathbf{DC}^T\right)^T = \mathbf{DC}^T, \left(\mathbf{A}^T \mathbf{C}\right)^T = \mathbf{A}^T \mathbf{C}, \left(\mathbf{D}^T \mathbf{B}\right)^T = \mathbf{D}^T \mathbf{B} \tag{A3}$$

and that

$$\mathbf{AD}^T - \mathbf{BC}^T = \mathbf{I}; \quad \mathbf{A}^T \mathbf{D} - \mathbf{C}^T \mathbf{B} = \mathbf{I}. \tag{A4}$$

In the case of uncoupled motion, all four matrices $\mathbf{A}, \mathbf{B}, \mathbf{C}, \mathbf{D}$ become diagonal making conditions (A3) trivial and turning (A4) into simple conditions for diagonal components:

$$A_{ii}D_{ii} - B_{ii}C_{ii} = 1; \quad i = 1,..,n \tag{A5}$$

equivalent to unity of determinants for individual *2x2* matrices in notations (A2):

$$M_i = \begin{bmatrix} A_{ii} & B_{ii} \\ C_{ii} & D_{ii} \end{bmatrix}; \quad \det M_i = 1. \tag{A6}$$

Before ending this Appendix, we would like to point out that one can use time as independent variable with traditional set of Canonical pairs:

$$\{x, P_x\}, \{y, P_y\}, \{z, P_z\} X^T = [\vec{r}, \vec{P}]. \tag{A7}$$

with traditional Hamiltonian [79]. The motion can be also expanded about the reference trajectory:

$$\vec{P} = \hat{x}P_x + \hat{y}P_y + \hat{z}(P_z + P_o); \quad \vec{q} = \vec{r} - \hat{z} \cdot z_o(t) = \hat{x}x + \hat{y}y + \hat{z}\zeta;$$

$$\left\{x, P_x = p_x + \frac{e}{c}A_x(\vec{r},t)\right\}; \left\{y, P_y = p_y + \frac{e}{c}A_y(\vec{r},t)\right\};$$

$$\left\{\zeta = z - z_o(t), P_z = p_z + \frac{e}{c}A_z(\vec{r},t) - P_o(t)\right\}; \quad P_o(t) = p_o(t) + \frac{e}{c}A_z(\vec{r}_o(t),t); \tag{A8}$$

$$q \equiv \{q_1, q_2, q_3\} = \{x, y, \zeta\}; \quad p \equiv \{p_1, p_2, p_3\} = \{\pi_x, \pi_y, \pi_z\};$$

$$H = H_o(q, p, t) = c\sqrt{1 + \left(\vec{P} - \frac{e}{c}\vec{A}\right)^2} + e\varphi - \dot{z}_o P_z - \zeta \dot{p}_o(t).$$

where $\vec{r}_o(t) = \hat{z} \cdot z_o(t)$ is the position of the reference particle (usually the center of the bunch) moving with designed momentum $p_o(t)$ along z-axis, and $\vec{A}(\vec{r},t), \varphi(\vec{r},t)$ are the vector and the scalar potential of the EM field. The reference particle has the design energy and velocity along the z-axis:

$$\mathbf{E}_o(t) \equiv \gamma_o(t)mc^2 = c\sqrt{p_o^2(t) + m^2c^2}; \quad \beta_o(t) \equiv \frac{\mathrm{v}_o(t)}{c} = \sqrt{1 - \gamma_o^{-2}(t)}. \tag{A9}$$

Since both systems are Hamiltonian and pairs have the same dimensionally, the transformation from *t*-based to *s*-based description represents Canonical transformation [79] and, therefore, both direct and inverse transformation are symplectic:

$$\tilde{X} = \tilde{X}(X,s) \Leftrightarrow X = X(\tilde{X},t);$$

$$\tilde{J}(X,s) = \frac{D\tilde{X}}{DX} \equiv \left[\frac{\partial \tilde{x}_i}{\partial x_j}\right]; \quad \tilde{J}^T S \tilde{J} = S; \quad J(\tilde{X},t) = \frac{DX}{D\tilde{X}} \equiv \left[\frac{\partial x_i}{\partial \tilde{x}_j}\right]; \quad J^T S J = S; \tag{A10}$$

$$S = \begin{bmatrix} \sigma & 0 & 0 \\ 0 & \sigma & 0 \\ 0 & 0 & \sigma \end{bmatrix}; \quad \sigma = \begin{bmatrix} 0 & 1 \\ -1 & 0 \end{bmatrix}; \quad J(\tilde{X}(X,s), t(s)) = \tilde{J}(X,s)^{-1};$$

In the case of linear matrix transformations used for infinitesimally small deviations from the reference there is direct connection between two transport matrices:

$$\delta X(t) = T(0|t)\delta X(0); \delta \tilde{X}(s) = M(0|s)\delta \tilde{X}(0);$$
$$T(0|t_o(s)) = \tilde{J}(s)^{-1} M(0|s)\tilde{J}(0);$$
(A11)

which makes these descriptions interchangeable.

Matrix $\mathbf{A}^{-1}\mathbf{B}$ plays critical role in the instability integral equation (43). It's properties can be studied for a Hamiltonian system describing a generic linear system:

$$H = \frac{1}{2}\xi^T \mathbf{H}(s)\xi; \ \mathbf{H}^T = \mathbf{H} = \begin{bmatrix} \mathbf{H}_q & \mathbf{H}_m^T \\ \mathbf{H}_m & \mathbf{H}_p \end{bmatrix}; \mathbf{H}_{q,p}^T = \mathbf{H}_{q,p}.$$
(A12)

Using equations of motion:

$$\mathbf{M}' = \mathbf{SH} \cdot \mathbf{M}; \ \mathbf{A}' = \mathbf{H}_m \mathbf{A} + \mathbf{H}_p \mathbf{C}; \ \mathbf{B}' = \mathbf{H}_m \mathbf{B} + \mathbf{H}_p \mathbf{D}.$$
(A13)

Taking into account that $(\mathbf{A}^{-1})' = -\mathbf{A}^{-1}\mathbf{A}'\mathbf{A}^{-1}$, we get

$$(\mathbf{A}^{-1}\mathbf{B})' = \mathbf{A}^{-1}\mathbf{B}' - \mathbf{A}^{-1}\mathbf{A}'\mathbf{A}^{-1}\mathbf{B} = \mathbf{A}^{-1}\mathbf{H}_p (\mathbf{D} - \mathbf{C}\mathbf{A}^{-1}\mathbf{B}),$$
(A14)

which can be turned into

$$(\mathbf{A}^{-1}\mathbf{B})' = \mathbf{A}^{-1}\mathbf{H}_p (\mathbf{A}^T)^{-1}$$
(A15)

$$\mathbf{D} - \mathbf{C}\mathbf{A}^{-1}\mathbf{B} = (\mathbf{D}\mathbf{A}^T - \mathbf{C}\mathbf{B}^T)(\mathbf{A}^{-1})^T = (\mathbf{A}^{-1})^T;$$
$$(\mathbf{A}^{-1}\mathbf{B})' = \mathbf{A}^{-1}\mathbf{H}_p (\mathbf{A}^{-1})^T$$
(A15)

using symplecticity conditions (A4) $\mathbf{A}^{-1}\mathbf{B} = \mathbf{B}^T(\mathbf{A}^T)^{-1}$ and $\mathbf{D}\mathbf{A}^T - \mathbf{C}\mathbf{B}^T = \mathbf{I}$ to show that

$$\mathbf{D} - \mathbf{C}\mathbf{A}^{-1}\mathbf{B} = \mathbf{D} - \mathbf{C}\mathbf{B}^T(\mathbf{A}^T)^{-1} = (\mathbf{D}\mathbf{A}^T - \mathbf{C}\mathbf{B}^T)(\mathbf{A}^T)^{-1} = (\mathbf{A}^T)^{-1}.$$

It is possible to show for an arbitrary accelerator [76,80-82] that $\mathbf{H}_p$ is a diagonal with positive diagonal terms:

$$\mathbf{H}_p = \frac{1}{\gamma_o \beta_o mc}\begin{bmatrix} 1 & 0 & 0 \\ 0 & 1 & 0 \\ 0 & 0 & (\gamma_o \beta_o)^{-2} \end{bmatrix}$$
(A16)

This allow us to prove that for an arbitrary accelerator with invertible matrix $\mathbf{A}$, the convolution $u(s)$ in eq. (43) is nonnegative monotonically growing function with $u'(s) > 0$.

$$u(s) = v_o^T \left( \mathbf{A}(s)^{-1} \mathbf{B}(s) \right) v_o \equiv \vec{v}_o \overrightarrow{\left( \mathbf{A}^{-1} \mathbf{B} \right)} \vec{v}_o; \; \mathbf{B}(0) = \mathbf{0} \rightarrow u(0) = 0; \; v(s) = \mathbf{A}^T(s)^{-1} v_o;$$

$$u'(s) = v_o^T \left( \mathbf{A}^{-1} \mathbf{B} \right)' v_o = v^T(s) \mathbf{H}_p(s) v(s) = \sum_{i=1}^{3} \mathbf{H}_{ii}(s) v_i^2(s) > 0. \tag{A16}$$

$$u(s) = \int_0^s \left( \sum_{i=1}^{3} \mathbf{H}_{ii}(s) v_i^2(s) \right) d\zeta \geq 0.$$

In other words, we proved that convolution of any constant vector with matrix $\mathbf{A}^{-1}\mathbf{B}$ is nonnegative monotonically growing function.

In the chase of uncoupled motion, all 2x2 matrices are diagonal and

$$\frac{d}{ds}\left( \mathbf{A}^{-1} \mathbf{B} \right)' = \begin{bmatrix} \alpha_1(s) & 0 & 0 \\ 0 & \alpha_2(s) & 0 \\ 0 & 0 & \alpha_3(s) \end{bmatrix} = \frac{1}{\gamma_o \beta_o mc} \begin{bmatrix} a_{11}^{-2} & 0 & 0 \\ 0 & a_{22}^{-2} & 0 \\ 0 & 0 & (a_{33}\gamma_o \beta_o)^{-2} \end{bmatrix}; \alpha_i(s) > 0;$$

$$\mu_i(s) = \int_0^s \alpha_i(\zeta) d\zeta \geq 0; \; \mathbf{A}^{-1} \mathbf{B} = \int_0^s \left( \mathbf{A}^{-1} \mathbf{B} \right)' d\zeta = \begin{bmatrix} \mu_1(s) & 0 & 0 \\ 0 & \mu_2(s) & 0 \\ 0 & 0 & \mu_3(s) \end{bmatrix} = \delta_{ij}\mu_i.$$

(A19)

i.e. diagonal term of matrix $\mathbf{A}^{-1}\mathbf{B}$ are monotonously growing positive functions:

$$\forall s_1 > s_2; \mu_i(s_1) > \mu_i(s_2). \tag{A20}$$

### Appendix B – Conditions for applicability of short period perturbations or, in other words, assumption of homogeneous infinite plasma.

Fourier or Laplace transformations are frequently used to solve the linearized Vlasov equation. The main problem from inhomogeneous distribution (or finite size of the beam) that it results in coupling Fourier harmonics of the perturbation with those the background, e.g., applying Fourier transformation to eq. (17)

$$\int dQ^3 e^{-i\vec{k}\vec{Q}} \left( \frac{\partial \breve{f}}{\partial t} + \frac{\partial F_o}{\partial \vec{Q}} \frac{\partial \breve{h}}{\partial \vec{P}} - \frac{\partial F_o}{\partial \vec{P}} \frac{\partial \breve{h}}{\partial \vec{Q}} \right) =$$

$$\frac{\partial \breve{f}_{\vec{k}}}{\partial t} + i \int d\kappa^3 \left\{ F_{o\vec{\kappa}} \left( \vec{\kappa} \cdot \frac{\partial \breve{h}_{\vec{k}-\vec{\kappa}}}{\partial \vec{P}} \right) - \breve{h}_{\vec{k}-\vec{\kappa}} \left( (\vec{k}-\vec{\kappa}) \cdot \frac{\partial F_{o\vec{\kappa}}}{\partial \vec{P}} \right) \right\},$$

does not result in separation of the Fourier harmonics. In this sense, this equation is as complicated the original Vlasov equation.

Let's consider a beam with typical scales of the inhomogeneity, $a_{x,y,z}$, which are not necessarily of the same order of magnitude:

$$\left|\frac{\partial F_o}{\partial x}\right| \propto \frac{F_o}{a_x}; \left|\frac{\partial F_o}{\partial y}\right| \propto \frac{F_o}{a_y}; \left|\frac{\partial F_o}{\partial z}\right| \propto \frac{F_o}{a_z}$$

The conditions for separation of the Fourier harmonics in (B.1) are easiest to derive in the comoving frame of reference, where $a_z$ is increased by relativistic factor $\gamma = E_o/mc^2$. For simplicity we will consider that motion of particles in the comoving frame is non-relativistic and we can neglect effects of magnetic field, e.g., assume $\vec{B} = 0$. In this case the perturbed Hamiltonian is a simple linear function of the electric field potential. In this case Maxwell equations are reduced to two simple equations for electric field:

$$div\vec{E} = 4\pi\rho; \ curl\vec{E} = 0. \tag{B1}$$

First, let's establish relations between parameters in laboratory and comoving frame. It is well known that that length of the bunch is scaled-up in the comoving frame by the relativistic factor $\gamma$

$$a_{z,cm} = \gamma a_z \tag{B2}$$

and the $\hat{k} = (\omega/c, \vec{k})$ transforms a 4-vector [87] proving well-known relations using the Lorentz transformation:

$$\vec{k}_{cm} \equiv \vec{\kappa} = \hat{z}\kappa_z + \vec{\kappa}_\perp; \omega_{cm} = 0; \ \vec{k}_{lab} = \hat{z}k_z + \vec{k}_\perp;$$

$$\vec{k}_\perp = \vec{\kappa}_\perp; \ k_z = \gamma_o\left(\kappa_z + \beta_o\frac{\omega_{cm}}{c}\right) = \gamma_o\kappa_z; \ \omega_{lab} = \gamma_o(\omega_{cm} + v_o\kappa_z) = v_o k_z; \tag{B3}$$

$$e^{i\vec{\kappa}\vec{r}_{cm}} \Leftrightarrow e^{i(\vec{k}\vec{r}_{lab} - v_o k_z t)},$$

where we use indexed "*cm*" and "*lab*" for the comoving and laboratory frames, correspondingly.

Further in this Appendix we will use the comoving frame and will drop the index "*c*". The natural conditions for neglecting the beam's edges, the transitions and the reflection effects is that there must be a significant number of oscillations in each direction at the typical scales of the inhomogeneity, e.g.:

$$\kappa_{x,y} a_{x,y} = k_{x,y} a_{x,y} \gg 2\pi; \ \gamma\kappa_z a_z = k_z a_z \gg 2\pi. \tag{B4}$$

Second, and much more convoluted, condition is that Fourier harmonic of the induced electric field (and therefor of the perturbation in the Hamiltonian) are linear functions of the harmonic of the charge density perturbation,

$$\rho_{\vec{\kappa}} = e\int_{-\infty}^{\infty} d\vec{r}e^{-i\vec{\kappa}\vec{r}} \int_{-\infty}^{\infty} \hat{f}_c(\vec{r},\vec{v},t)d\vec{v}. \tag{B5}$$

where $\hat{f}_c$ is perturbation of the distribution function in the comoving frame. In the infinite charged plasma periodic density perturbation results is periodic electric field aligned with the $\vec{\kappa}$-vector:

$$\vec{E} = \vec{E}_{\vec{\kappa}} e^{i\vec{\kappa}\vec{r}}; \vec{E}_{\vec{\kappa}} = \vec{E}_{\vec{\kappa}\|} + \vec{E}_{\vec{\kappa}\perp}; \vec{E}_{\vec{\kappa}\|} = \vec{\kappa}\frac{(\vec{\kappa} \cdot \vec{E}_{\vec{\kappa}})}{\vec{\kappa}^2} = \frac{\vec{\kappa}}{\kappa} E_{\kappa};$$

$$curl\vec{E} = 0 \to \vec{\kappa} \times \vec{E}_{\vec{\kappa}} = 0 \to \vec{E}_{\vec{\kappa}\perp} = 0 \to \vec{E} = \frac{\vec{\kappa}}{\kappa} E_{\kappa}; \quad (B6)$$

$$div\vec{E} = 4\pi\rho_{\vec{\kappa}} e^{i\vec{\kappa}\vec{r}} \to i\vec{\kappa} \cdot \vec{E}_{\vec{\kappa}} = i\kappa E_{\kappa} = 4\pi\rho_{\vec{\kappa}},$$

resulting in well known

$$\vec{E}_{\vec{\kappa}} \cong 4\pi\rho_{\vec{\kappa}} \frac{\vec{\kappa}}{i\vec{\kappa}^2}. \quad (B7)$$

For the non-uniform density resulting field will deviate from intuitive extensions of (B7) by $\delta\vec{E}$:

$$\vec{E} = 4\pi\rho_{\vec{\kappa}}(\vec{r})\frac{\vec{\kappa}}{i\vec{\kappa}^2} e^{i\vec{\kappa}\vec{r}} + \delta\vec{E}, \quad (B8)$$

and we need to find conditions when $\kappa|\delta\vec{E}| \ll 4\pi|\rho_{\vec{\kappa}}|$. We get the following using (B3):

$$div\vec{E} = 4\pi\rho_{\vec{\kappa}}(\vec{r})e^{i\vec{\kappa}\vec{r}} + 4\pi\frac{(\vec{\kappa} \cdot \vec{\nabla}\rho_{\vec{\kappa}}(\vec{r}))}{\vec{\kappa}^2} e^{i\vec{\kappa}\vec{r}} + div\delta\vec{E} = 4\pi\rho_{\vec{\kappa}}(\vec{r})e^{i\vec{\kappa}\vec{r}};$$

$$curl\vec{E} = 4\pi\frac{(\vec{\kappa} \times \vec{\nabla}\rho_{\vec{\kappa}}(\vec{r}))}{\vec{\kappa}^2} e^{i\vec{\kappa}\vec{r}} + curl\delta\vec{E} = 0;$$

$$div\delta\vec{E} = -4\pi\frac{(\vec{\kappa} \cdot \vec{\nabla}\rho_{\vec{\kappa}}(\vec{r}))}{\vec{\kappa}^2} e^{i\vec{\kappa}\vec{r}} \sim \kappa|\delta\vec{E}| \quad (B9)$$

$$curl\delta\vec{E} = -4\pi\frac{(\vec{\kappa} \times \vec{\nabla}\rho_{\vec{\kappa}}(\vec{r}))}{\vec{\kappa}^2} e^{i\vec{\kappa}\vec{r}} \sim \kappa|\delta\vec{E}|$$

While the error estimation resulting from $div\vec{E} = 4\pi\rho_{\vec{\kappa}}$ improves on the intuitive requirement (B4):

$$\frac{\partial\delta E_x}{\partial x} + \frac{\partial\delta E_y}{\partial y} + \frac{\partial\delta E_z}{\partial z} = -4\pi\frac{e^{i\vec{\kappa}\vec{r}}}{\vec{\kappa}^2}\left(\kappa_x \cdot \frac{\partial\rho_{\vec{\kappa}}}{\partial x} + \kappa_y \cdot \frac{\partial\rho_{\vec{\kappa}}}{\partial y} + \kappa_z \cdot \frac{\partial\rho_{\vec{\kappa}}}{\partial y}\right); \left|\frac{\partial\rho_{\vec{\kappa}}}{\partial x_i}\right| \sim \frac{|\rho_{\vec{\kappa}}|}{\sigma_i};$$

$$\left|-ie^{i\vec{\kappa}\vec{r}}\left(\frac{\partial\delta E_x}{\partial x} + \frac{\partial\delta E_y}{\partial y} + \frac{\partial\delta E_z}{\partial z}\right)\right| \sim \left(|\kappa_x \delta E_x| + |\kappa_y \delta E_y| + |\kappa_z \delta E_y|\right) \sim 4\pi\frac{|\rho_{\vec{\kappa}}|}{k^2}\left(\frac{|\kappa_x|}{a_x} + \frac{|\kappa_y|}{a_y} + \frac{|\kappa_z|}{\gamma a_z}\right) \quad (B10)$$

$$\left|\frac{div\delta\vec{E}}{div\vec{E}}\right| \sim \frac{\frac{|\kappa_x|}{a_x} + \frac{|\kappa_y|}{a_y} + \frac{|\kappa_z|}{\gamma a_z}}{k^2} \ll 1$$

the error estimations resulting from $curl\vec{E} = 0$

$$\mathrm{curl}\,\delta\vec{E} = -4\pi \frac{\left(\vec{\kappa}\times\vec{\nabla}\rho_{\vec{\kappa}}(\vec{r})\right)}{\vec{\kappa}^2} e^{i\vec{\kappa}\vec{r}} = \frac{4\pi}{\vec{\kappa}^2} e^{i\vec{\kappa}\vec{r}} \times$$

$$\left\{ \hat{x}\left(\kappa_y \frac{\partial \rho_{\vec{\kappa}}}{\partial z} - \kappa_z \frac{\partial \rho_{\vec{\kappa}}}{\partial y}\right) + \hat{y}\left(\kappa_z \frac{\partial \rho_{\vec{\kappa}}}{\partial x} - \kappa_x \frac{\partial \rho_{\vec{\kappa}}}{\partial z}\right) + \hat{z}\left(\kappa_x \frac{\partial \rho_{\vec{\kappa}}}{\partial y} - \kappa_y \frac{\partial \rho_{\vec{\kappa}}}{\partial x}\right) \right\}$$

$$\frac{\partial \delta E_z}{\partial y} - \frac{\partial \delta E_y}{\partial z} = \frac{4\pi}{\vec{\kappa}^2} e^{i\vec{\kappa}\vec{r}} \left(\kappa_y \frac{\partial \rho_{\vec{\kappa}}}{\partial z} - \kappa_z \frac{\partial \rho_{\vec{\kappa}}}{\partial y}\right); \quad \frac{\partial \delta E_x}{\partial z} - \frac{\partial \delta E_z}{\partial x} = \frac{4\pi}{\vec{\kappa}^2} e^{i\vec{\kappa}\vec{r}} \left(\kappa_z \frac{\partial \rho_{\vec{\kappa}}}{\partial x} - \kappa_x \frac{\partial \rho_{\vec{\kappa}}}{\partial z}\right); \quad (B11)$$

$$\frac{\partial \delta E_y}{\partial x} - \frac{\partial \delta E_x}{\partial y} = \frac{4\pi}{\vec{\kappa}^2} e^{i\vec{\kappa}\vec{r}} \left(\kappa_x \frac{\partial \rho_{\vec{\kappa}}}{\partial y} - \kappa_y \frac{\partial \rho_{\vec{\kappa}}}{\partial x}\right);$$

is much more important because it links all three dimensions:

$$\left|\frac{\partial \delta E_z}{\partial y}\right| + \left|\frac{\partial \delta E_y}{\partial z}\right| \sim \left|\kappa_y \delta E_z\right| + \left|\kappa_z \delta E_y\right| \sim \frac{4\pi}{\kappa^2} |\rho_{\vec{\kappa}}| \left(\frac{|\kappa_y|}{\gamma a_z} + \frac{|\kappa_z|}{a_y}\right);$$

$$\left|\frac{\partial \delta E_x}{\partial z}\right| + \left|\frac{\partial \delta E_z}{\partial x}\right| \sim \left|\kappa_z \delta E_x\right| + \left|\kappa_x \delta E_z\right| \sim \frac{4\pi}{k^2} |\rho_{\vec{\kappa}}| \left(\frac{|\kappa_x|}{\gamma a_z} + \frac{|\kappa_z|}{a_x}\right); \quad (B12)$$

$$\left|\frac{\partial \delta E_x}{\partial y}\right| + \left|\frac{\partial \delta E_y}{\partial x}\right| \sim \left|\kappa_y \delta E_x\right| + \left|\kappa_x \delta E_y\right| \sim \frac{4\pi}{k^2} |\rho_{\vec{\kappa}}| \left(\frac{|\kappa_x|}{a_y} + \frac{|\kappa_y|}{a_x}\right).$$

This allows us to estimate errors for each component of electric field:

$$|\vec{E}| \cong 4\pi \frac{|\rho_{\vec{\kappa}}|}{\kappa}; \quad |\delta E_x| \sim |\vec{E}|\left(\frac{1}{\kappa a_x} + \frac{1}{\kappa a_y}\frac{|\kappa_y|}{|\kappa_x|} + \frac{1}{\gamma \kappa a_z}\frac{|\kappa_z|}{|\kappa_x|}\right);$$

$$|\delta E_y| \sim |\vec{E}|\left(\frac{1}{\kappa a_x}\frac{|\kappa_x|}{|\kappa_y|} + \frac{1}{\kappa a_y} + \frac{1}{\gamma \kappa a_z}\frac{|\kappa_z|}{|\kappa_y|}\right); \quad |\delta E_z| \sim |\vec{E}|\left(\frac{1}{\kappa a_x}\frac{|\kappa_x|}{|\kappa_z|} + \frac{1}{\kappa a_y}\frac{|\kappa_y|}{|\kappa_z|} + \frac{1}{\gamma \kappa a_z}\right); \quad (B13)$$

and

$$|\delta E_y| \sim |\vec{E}|\left(\frac{1}{\gamma \kappa a_z}\frac{|\kappa_y|}{|\kappa_z|} + \frac{1}{\kappa a_y}\right); \quad |\delta E_y| \sim |\vec{E}|\left(\frac{1}{\kappa a_y} + \frac{1}{\kappa a_x}\frac{|\kappa_y|}{|\kappa_x|}\right);$$

$$|\delta E_z| \sim |\vec{E}|\left(\frac{1}{\gamma \kappa a_z} + \frac{1}{\kappa a_y}\frac{|\kappa_z|}{|\kappa_y|}\right); \quad |\delta E_z| \sim |\vec{E}|\left(\frac{1}{\gamma \kappa a_z} + \frac{1}{\kappa a_x}\frac{|\kappa_z|}{|\kappa_x|}\right); \quad (B14)$$

$$|\delta E_x| \sim |\vec{E}|\left(\frac{1}{\gamma \kappa a_z}\frac{|\kappa_x|}{|\kappa_z|} + \frac{1}{\kappa a_x}\right); \quad |\delta E_x| \sim |\vec{E}|\left(\frac{1}{\kappa a_x} + \frac{1}{\kappa a_y}\frac{|\kappa_x|}{|\kappa_y|}\right).$$

Now, let's introduce the following definitions:

$$\varepsilon_x = \frac{1}{\kappa a_x}; \varepsilon_y = \frac{1}{\kappa a_y}; \varepsilon_z = \frac{1}{\gamma \kappa a_z};$$

And

$$r_{xy} = \frac{|\kappa_x|}{|\kappa_y|}; r_{xz} = \frac{|\kappa_x|}{|\kappa_z|}; r_{yz} = \frac{|\kappa_y|}{|\kappa_z|};$$

(B15)

and rewrite (B13-14) as

$$|\delta E_x| \sim |\vec{E}| \cdot \min\left(\varepsilon_x + \frac{\varepsilon_y}{r_{xy}} + \frac{\varepsilon_z}{r_{xz}}\right); |\delta E_y| \sim |\vec{E}|\left(\varepsilon_x r_{xy} + \varepsilon_y + \frac{\varepsilon_z}{r_{yz}}\right); |\delta E_z| \sim |\vec{E}|\left(\varepsilon_x r_{xz} + \varepsilon_y r_{yz} + \varepsilon_z\right);$$

$$|\delta E_y| \sim |\vec{E}|(\varepsilon_z r_{yz} + \varepsilon_y); |\delta E_y| \sim |\vec{E}|\left(\varepsilon_y + \frac{\varepsilon_x}{r_{xy}}\right); |\delta E_z| \sim |\vec{E}|\left(\varepsilon_z + \frac{\varepsilon_y}{r_{yz}}\right); |\delta E_z| \sim |\vec{E}|\left(\varepsilon_z + \frac{\varepsilon_x}{r_{xz}}\right);$$ (B16)

$$|\delta E_x| \sim |\vec{E}|(\varepsilon_z r_{xz} + \varepsilon_x); |\delta E_x| \sim |\vec{E}|(\varepsilon_x + \varepsilon_y r_{xy}).$$

And finally, the combination gives of all estimations result in the following:

$$\min\left(r, \frac{1}{r}\right) = 1, r \geq 0$$

$$|\delta E_x| \sim |\vec{E}| \cdot \min\left(\varepsilon_x + \frac{\varepsilon_y}{r_{xy}} + \frac{\varepsilon_z}{r_{xz}}, \varepsilon_x + \varepsilon_z r_{xz}, \varepsilon_x + \varepsilon_y r_{xy}\right) \leq \varepsilon_x + \varepsilon_y + \varepsilon_z;$$

$$|\delta E_y| \sim |\vec{E}| \min\left(\varepsilon_x r_{xy} + \varepsilon_y + \frac{\varepsilon_z}{r_{yz}}, \varepsilon_y + \varepsilon_z r_{yz}, \frac{\varepsilon_x}{r_{xy}} + \varepsilon_y\right) \leq \varepsilon_x + \varepsilon_y + \varepsilon_z;$$ (B17)

$$|\delta E_z| \sim |\vec{E}| \min\left(\varepsilon_x r_{xz} + \varepsilon_y r_{yz} + \varepsilon_z, \frac{\varepsilon_y}{r_{yz}} + \varepsilon_z, \frac{\varepsilon_x}{r_{xz}} + \varepsilon_z\right) \leq \varepsilon_x + \varepsilon_y + \varepsilon_z;$$

It means that

$$\kappa a_x \gg 1; \kappa a_y \gg 1; \gamma \kappa a_z \gg 1.$$ (B18)

are sufficient conditions in the co-moving frame for eq. (B7) to be a valid approximation for the electric field.

Lorentz transformation (B#) changes these conditions to the lab-frame as follows:

$$a_{x,y} \cdot \sqrt{\vec{k}_\perp^2 + \frac{\vec{k}_z^2}{\gamma^2}} \gg 1;\ a_z \cdot \sqrt{\gamma^2 \vec{k}_\perp^2 + \vec{k}_z^2} \gg 1.$$ (B19)

These conditions are most important for the case of the longitudinal density modulation

$$\vec{k}_{lab} = \hat{z} k_{//};\ k_{//} \gg \max\left(\frac{\gamma}{a_{x,y}}, \frac{1}{a_z}\right),$$ (B20)

which we will use late in this paper.

**Appendix C. Fields transfer from the comoving frame**

An instantaneous comoving frame (with a fixed velocity $\mathbf{v_o}$ along the direction of the reference trajectory) can be used for EM field evaluation and transferring it back to laboratory frame.

In this Appendix we will use metric of special relativity for 4-dimetnetional time and space [79]:

$$a^i = (a_o, \vec{a}); b_i = (b_o, -\vec{b}); a^i b_i = a_o b_o - \vec{a} \cdot \vec{b}, \tag{C1}$$

three 4-vectors: the time space, the k-vector and the 4-potentail of EM field,

$$x^i = (ct, \vec{r}); \; k^i = \left(\frac{\omega}{c}, \vec{k}\right); \; A^i = (\varphi, \vec{A}), \tag{C2}$$

and fact that phase of oscillations is 4-scalar

$$\phi = -k_i x^i = \vec{k}\vec{r} - \omega t = \text{inv}, \tag{C3}$$

and is invariant of Lorentz transformation. Here we will use Lorentz transformation into and from inertial frame moving along z-axis with velocity $\mathbf{v}_o$:

$$\hat{z} = \vec{\tau}(s_o); \; \vec{v}_o = v_o \cdot \hat{z}(s_o) = \text{const}; \beta_o = \frac{v_o}{c}; \gamma_o = \frac{1}{\sqrt{1-\beta_o^2}},$$

$$a_{ol} = \gamma_o(a_{oc} + \beta_o a_{zc}); a_{lz} = \gamma_o(a_{zc} + \beta_o a_{oc});$$
$$a_{oc} = \gamma_o(a_{ol} - \beta_o a_{zl}); a_{zc} = \gamma_o(a_{zl} - \beta_o a_{ol}); \tag{C4}$$
$$a_{(x,y)l} = a_{(x,y)c} = a_{(x,y)l} = a_{x,y};$$

where we use subscripts $c$ and $l$ for variables in the comoving and the laboratory frame, correspondingly.

Using standard assumption that motion of particles in this frame is non-relativistic, $|\vec{v}_c| \ll c$, allows us to neglect magnetic field and use zero vector potential [14]

$$\varphi_c^k = (\varphi_c, \vec{A}_c); \; \vec{A}_c = 0; \; \Delta\varphi_c = -4\pi\rho(\vec{r}_c). \tag{C5}$$

Applying Fourier transform $f(\vec{k}_c) = \int f(\vec{r}_c) e^{-i\vec{k}_c \vec{r}_c} d\vec{r}_c^3$, we obtain

---

[14] Here we are using standard assumption that motion of particles in the comoving frame is non-relativistic and magnetic field can be neglected. Second standard assumption is that plasma frequency is slow when compared with $c|\vec{\kappa}|$: $\omega_p \ll c|\vec{\kappa}|$ allows to neglect second driving term in Maxwell equation from magnetic field: $\text{curl}\vec{B} = 4\pi\vec{j} + \frac{1}{c}\frac{\partial \vec{E}}{\partial t}$.

$$\varphi_c(\vec{k}_c) = \frac{4\pi\rho_{c\vec{k}}}{\vec{k}_c^2}; \vec{A}_c(\vec{k}_c) = 0; \tilde{\varphi}_c = \frac{4\pi\rho_{c\vec{k}}}{\vec{k}_c^2} e^{i\vec{k}_c \vec{r}_c} \equiv \frac{4\pi\rho_{c\vec{k}}}{\vec{k}_c^2} e^{-ik^i x_i};$$
$$\tilde{A}_c^i = (\tilde{\varphi}_c, 0, 0, 0);$$
(C6)

where we used the facts that $\omega_c = 0$. The 4-potential (C6) we can be easily transferred to the laboratory frame:

$$\tilde{\varphi}_l = \gamma_o \tilde{\varphi}_c; \vec{\tilde{A}}_l = \vec{\tau} \cdot \gamma_o \beta_o \tilde{\varphi}_c; \tilde{A}_l^i = \gamma_o \tilde{\varphi}_c \cdot (1, \beta_o, 0, 0).$$
(C7)

and expressed through the components in the laboratory frame:

$$\tilde{\varphi}_l = -\frac{4\pi\rho_{\vec{k}l}}{\vec{k}_l^2 - \beta_o^2 k_{zl}^2} e^{i(\vec{k}_l \vec{r}_l - k_{zl}\beta_o ct_l)}; \vec{\tilde{A}}_l = \hat{z}\beta_o \tilde{\varphi}_l; \hat{z} = \vec{\tau}(s_o)$$
(C8)

taking into account scaling of the density by the factor $\gamma_o$:

$$\rho_{\vec{k}l} = \gamma_o \rho_{\vec{k}c},$$
(C9)

and confection between k-vectors:

$$k_{zl} = \gamma_o k_{zc}; \omega_l = ck_{ol} = \gamma_o \beta_o ck_{zc} = \beta_o ck_{zl}; \vec{k}_{\perp l} = \vec{k}_{\perp s};$$
$$\vec{k}_c^2 = k_{zc}^2 + \vec{k}_{\perp c}^2 = \vec{k}_l^2 - \beta_o^2 k_{zl}^2.$$
(C10)

Equation (C8) is identical to the EM potentials in Cartesian coordinates, that we derived in the next Appendix.

**Appendix D. Expression for charge and current density modulation**

Solving Maxwell equations require knowledge of the charge and current densities as functions of coordinates and time:

$$div\vec{B} = 0; div\vec{E} = 4\pi\rho; curl\vec{E} = -\frac{1}{c}\frac{\partial \vec{B}}{\partial t}; curl\vec{B} = \frac{1}{c}\frac{\partial \vec{E}}{\partial t} + \frac{4\pi}{c}\vec{j};$$
$$\rho(\vec{r},t) = e\int f(\vec{r},\vec{v},t)d\vec{v}^3; \vec{j}(\vec{r},t) = e\int \vec{v}f(\vec{r},\vec{v},t)d\vec{v}^3,$$
(D1)

where $f(\vec{r},\vec{v},t)$ is the particles distribution function in the $(\vec{r},\vec{v})$ configuration space. Using $s$ as independent variable makes connection between $\rho, \vec{j}$ and the phase space distribution function $F(q,p,s)$ non-trivial, where $(q,p)$ is conjugate Canonical set of coordinates and momenta. This Appendix is dedicated for establishing such connection and finding corresponding 4-potentila of the EM field.

Let's introduce instantaneous Cartesian coordinate system with z-axis along the reference trajectory at $s = s_o$ (see equation (6) in the main text):

$$\hat{x} = \vec{n}(s_o); \hat{y} = \vec{b}(s_o); \hat{z} = \vec{\tau}(s_o) = \frac{d\vec{r}_o(s)}{ds}\bigg|_{s=s_o}; \frac{d\vec{n}}{ds} = K(s)\cdot\vec{\tau} - \kappa(s)\cdot\vec{b}; \frac{d\vec{b}}{ds} = \kappa(s)\cdot\vec{n};$$

$$\vec{r} = \vec{r}_o(s_o) + \hat{x}\cdot x + \hat{y}\cdot y + \hat{z}\cdot z \equiv \vec{r}_o(s) + \vec{n}(s)q_1 + \vec{b}(s)q_2;$$

$$q_1 = \vec{n}(s)(\hat{x}x + \hat{y}\cdot y + \vec{r}_o(s_o) - \vec{r}_o(s)); q_2 = \vec{b}(s)(\hat{x}x + \hat{y}\cdot y + \vec{r}_o(s_o) - \vec{r}_o(s));$$

$$q_3 = c(t_o(s) - t).$$

(D2)

with following ratios in the vicinity of $\vec{r}_o(s)$:

$$dx = dq_1 + q_2\kappa(s_o)ds; dy = dq_2 - q_1\kappa(s_o)ds; \quad dz = (1 + q_1 K(s_o))ds;$$

$$q_3 = c(t_o(s) - t); dq_3 = \frac{ds}{\beta_o(s_o)} - cdt.$$

(D3)

where we used eq. (6) for reference trajectory. At fixed s:

$$ds = 0 \rightarrow dz = 0; dq_1 = dx; dq_1 = dy; dq_3 = -cdt.$$

(D4)

Number of particles confined in infinitesimal volume $dq^3$ at fixed $s=s_o$ is defined as:

$$dn = |dq^3| \int \tilde{f}(q,p,s_o)dp^3 = cdxdydt \int \tilde{f}(q,p,s_o)dp^3$$

(D5)

is identical to number of particles passing though the elementary area $dxdy$ locates at $s=s_o$ in time interval $dt$:

$$\vec{j}(\vec{r},t) = \int \vec{v} f(\vec{r},\vec{v},t)d\vec{v}^3; \quad dn = d\vec{a}\cdot\vec{j}\cdot dt; d\vec{a} = \hat{z}dxdy;$$

$$dn = dxdydt \int v_z f(\vec{r},\vec{v},t)d\vec{v}^3,$$

(D6)

resulting in

$$\int \tilde{f}(q,p,s)dp^3 = \frac{1}{c}\int v_z f(\vec{r},\vec{v},t)d\vec{v}^3$$

(D7)

Using paraxial approximation $v_z = v_o + \delta v; |\delta v| \ll v_o$, and neglecting $\delta v$ in the integral (D7), we can express $\rho, \vec{j}$ using phases space distribution function as:

$$\rho(\vec{r},t) \cong \frac{\rho(q,s)}{\beta_o(s)}; \vec{j}(\vec{r},t) \cong \hat{z}c\rho(\vec{r},t); \rho(q,s) = e\int \tilde{f}(q,p,s)dp^3.$$

(D8)

Applying Fourier transformation[15]

$$g_{\vec{k}} \equiv g(k_1, k_2, k_3, s) = \int g(\vec{q},s)e^{-i\vec{k}\vec{q}}d\vec{q}^3$$

---

[15] In this Appendix, we use interchangeably both the compact, $g_{\vec{k}}$, and detailed, $g(k_1,k_2,k_3,s)$, notation for the Fourier components defined in the accelerator coordinates.

to (D8) we obtain expressions for $\rho, \vec{j}$ at fixed $s$:

$$\rho_{\vec{k}} = e\frac{\tilde{f}_{\vec{k}}}{\beta_o}; \quad \vec{j}_{\vec{k}} = e\hat{z}c\tilde{f}_{\vec{k}}; \quad \tilde{f}_{\vec{k}} \equiv \tilde{f}(\vec{k},s) = \int e^{-i\vec{k}\vec{q}}\tilde{f}(q,p,s)dp^3 dq^3;$$

$$\rho = \frac{1}{(2\pi)^3}\frac{1}{\beta_o}\int \tilde{f}_{\vec{k}}e^{i\vec{k}\vec{q}}d\vec{k}^3; \quad \vec{j} = \frac{e\hat{z}c}{(2\pi)^3}\int \tilde{f}_{\vec{k}}e^{i\vec{k}\vec{q}}d\vec{k}^3. \tag{D9}$$

Let's calculate Fourier harmonic of the density

$$\rho(\vec{\kappa},\omega) = \int \rho e^{-i(\vec{\kappa}\vec{r}-\omega t)}d\vec{r}^3 dt = \frac{e}{(2\pi)^3}\int d\vec{k}^3 \int \frac{\tilde{f}_{\vec{k}}(s)}{\beta_o(s)}e^{i(\vec{k}\vec{q}-\vec{\kappa}\vec{r}+\omega t)}d\vec{r}^3 dt. \tag{D10}$$

Using $q_1 = x; q_2 = y$ and combining terms in the exponent:

$$\vec{k}\vec{q} - \vec{\kappa}\vec{r} + \omega t = (k_1 - \kappa_x)x + (k_2 - \kappa_y)y - (\omega - k_3 c)t + k_3 ct_o(s) - \kappa_z z; \tag{D11}$$

making five integrals to be trivial:

$$\frac{1}{(2\pi)^3}\iint e^{i(k_1-\kappa_x)x}e^{i(k_2-\kappa_y)y}e^{-i(\omega-ck_3)t}dx\,dy\,dt = \delta(k_1-\kappa_x)\delta(k_2-\kappa_y)\delta(\omega-ck_3);$$

$$\iint \delta(k_1-\kappa_x)\delta(k_2-\kappa_y)\delta(\omega-c\kappa_y)\tilde{f}(k_1,k_2,k_3,s)dk^3 = \tilde{f}\left(\kappa_x,\kappa_y,\frac{\omega}{c},s\right); \tag{D12}$$

$$\rho(\vec{\kappa},\omega) = \frac{e}{c}\int \frac{\tilde{f}\left(\kappa_x,\kappa_y,\frac{\omega}{c},s\right)}{\beta_o(s)}e^{i(\omega t_o(s)-\kappa_z z)}dz.$$

At this point we can use our assumption that the scale of variation of the accelerator parameters (such as curvature, $\beta_o$, etc.) are much larger than scale of modulation, In addition, we assume that evolution of the density modulation is also much slower than fast oscillating term $e^{i\omega t_o(s)}$. This assumption will allow us to move $\tilde{f}$ and $\beta_o$ outside the integral and also to expand the arrival time of the reference particle with respect of azimuth $\bar{s}$, where we locate origin of z-axis:

$$t_o(s) \cong t_o(\bar{s}) + \frac{z}{c\beta_o(\bar{s})}, \tag{D13}$$

and arrive to the final relation between Fourier components in two systems of coordinates:

$$\rho(\vec{\kappa},\omega) = \frac{2\pi e}{c}\tilde{f}(\kappa_x,\kappa_y,\beta_o\kappa_z,\bar{s})\delta(k_o - \beta_o\kappa_z)e^{i\beta_o\kappa_z ct_o(\bar{s})}; \quad k_o = \frac{\omega}{c};$$

$$\vec{j}(\vec{\kappa},\omega) = \hat{z}c\beta_o\rho(\vec{\kappa},\omega) \tag{D14}$$

where we used $\delta\left(\frac{k_o}{\beta_o} - \kappa_z\right) = \beta_o \delta(k_o - \beta_o\kappa_z)$ and singularity of Dirac's δ-function:

$$g(x)\delta(x-y) = g(y)\delta(x-y).$$

To find 4-potential induced by such perturbation we can use Lorenz gauge $\frac{\partial \varphi}{c\partial t} + div\vec{A} = 0$ providing for separation of equations for each component of 4-potentail [88][16]:

$$\frac{\partial^2 \varphi}{c^2 \partial t^2} - \Delta\varphi = 4\pi\rho; \quad \frac{\partial^2 \vec{A}}{c^2 \partial t^2} - \Delta\vec{A} = 4\pi\vec{j},$$

which can be Fourier transformed to:

$$\varphi(\vec{\kappa},\omega)e^{i(\vec{\kappa}\vec{r}-\omega t)} = \frac{4\pi\rho(\vec{\kappa},\omega)}{\vec{\kappa}^2 - k_o^2}e^{i(\vec{\kappa}\vec{r}-\omega t)} = \frac{8\pi^2 e}{c}\frac{\tilde{f}(\kappa_x,\kappa_y,\beta_o\kappa_z,\bar{s})}{\vec{\kappa}^2 - \beta_o^2\kappa_z^2}\delta(k_o - \beta_o\kappa_z)e^{i(\vec{\kappa}\vec{r}+\omega(t_o(\bar{s})-t))};$$
(D15)

$$\vec{A}(\vec{\kappa},\omega) = \vec{z}\varphi(\vec{\kappa},\omega)\frac{k_o}{\kappa_z} = \vec{z}\beta_o\varphi(\vec{\kappa},\omega).$$

In the inverse Fourier transform

$$\varphi(\vec{r},t) = \frac{1}{(2\pi)^4}\int \varphi(\vec{\kappa},\omega)e^{i(\vec{\kappa}\vec{r}-\omega t)}d\omega d\vec{\kappa}^3,$$

δ-function makes integral over $\omega$ is straight forward:

$$\frac{1}{2\pi}\int e^{i\omega(t_o(\bar{s})-t)}g\left(\vec{\kappa},\frac{\omega}{c}\right)\delta\left(\frac{\omega}{c} - \beta_o\kappa_z\right)d\omega = \frac{c}{2\pi}g(\vec{\kappa},\beta_o\kappa_z)e^{i\beta_o\kappa_z\tau(\bar{s})}; \tau(\bar{s}) = c(t_o(\bar{s})-t), \quad \text{(D16)}$$

with the remaining integral of

$$\varphi(\vec{r},t) = 4\pi e \int \frac{\tilde{f}(\kappa_x,\kappa_y,\beta_o\kappa_z,\bar{s})}{\vec{\kappa}^2 - \beta_o^2\kappa_z^2}e^{i(\vec{\kappa}\vec{r}+\beta_o\kappa_z\tau(\bar{s}))}\frac{d\vec{\kappa}^3}{(2\pi)^3}; \quad \vec{A}(\vec{r},t) = \vec{z}\beta_o\varphi(\vec{r},t). \quad \text{(D17)}$$

Taking into account expansion (D13), the exponent in (D17) can be expressed using the accelerator coordinates:

$$\vec{\kappa}\vec{r} + \beta_o\kappa_z\tau(\bar{s}) = \vec{k}\vec{q}; k_{1,2} = \kappa_{x,y}; k_3 = \beta_o\kappa_z; \quad \text{(D18)}$$

and using ratio $\beta_o d\vec{\kappa}^3 = d\vec{k}^3$ we get expression connecting the 4-potentail and density perturbation in the accelerator coordinates:

$$\varphi(\vec{q},s) \equiv \varphi(\vec{r},t) = 4\pi e \beta_o \gamma_o^2 \int \frac{\tilde{f}_{\vec{k}}}{\gamma_o^2\beta_o^2\vec{k}_\perp^2 + k_z^2}e^{i\vec{k}\vec{q}}\frac{d\vec{k}^3}{(2\pi)^3}; \vec{A}(\vec{q},s) = \vec{z}\beta_o\varphi(\vec{q},s), \quad \text{(D19)}$$

where we take into account

---

[16] The Lorentz gauge can be used for time-dependent component of the EM field, which is of interest in this paper.

**Appendix E. Perturbed Hamiltonian**

As derived in Appendix D, density perturbation results in additional 4-potentail

$$\delta\varphi^i = \{\delta\varphi, \delta\vec{A}\}; \delta\vec{A} = \hat{z}\beta_o \delta\varphi. \tag{E1}$$

which we will consider being infinitesimally small: $\delta\varphi \sim O(\varepsilon), \varepsilon \ll 1$. Goal of this Appendix is to define additional term of the reduced accelerator Hamiltonian (7-9) resulting from the density perturbation:

$$\begin{aligned}
h^* = &-(1+Kq_1)\sqrt{\frac{(E_o + cP_3 - e\varphi_\perp - e\delta\varphi)^2}{c^2} - m^2c^2 - \left(P_1 - \frac{e}{c}A_1\right)^2 - \left(P_2 - \frac{e}{c}A_2\right)^2} \\
&-\frac{e}{c}(1+Kq_1)(A_z + \beta_o\delta\varphi) + \kappa q_1 P_2 - \kappa q_2 P_1 - \frac{c}{v_o(s)}P_3 + q_3 \frac{d}{ds}\left(E_o(s) + e\frac{\varphi(\vec{r}_o(s),t)}{c}\right)
\end{aligned} \tag{E2}$$

where we used explicit expression for $A_3$ component of the vector potential (7). Perturbation of the Hamiltonian is coming only from first two terms in r.h.s of (E2).

$$\delta h^* = -(1+Kq_1)\left(\sqrt{\frac{(E - e\delta\varphi)^2}{c^2} - m^2c^2 - \vec{p}_\perp^2} - p_z + \frac{e}{c}\beta_o\delta\varphi\right); \quad p_z = \sqrt{\frac{E^2}{c^2} - m^2c^2 - \vec{p}_\perp^2};$$

$$\sqrt{\frac{(E - e\delta\varphi)^2}{c^2} - m^2c^2 - \vec{p}_\perp^2} - p_z = -\frac{E}{cp_z}\frac{e}{c}\delta\varphi + O(\delta\varphi)^2; \tag{E3}$$

$$E = H - e\varphi; \quad \beta_o = \frac{v_o}{c}; \beta_z = \frac{v_z}{c} = \frac{cp_z}{E}; 1 - \beta_o^2 = \gamma_o^{-2};$$

$$\delta h^* = (1+Kq_1)\frac{e}{c}\delta\varphi\left(\frac{c}{v_z} - \frac{v_o}{c}\right) = \frac{1}{\beta_o \gamma_o^2}\frac{e}{c}\delta\varphi\left\{1 + \gamma_o^2\left(\frac{\beta_o}{\beta_z} - 1\right)\right\}(1+Kq_1).$$

First, in paraxial approximation term $|Kq_1| \ll 1$ can be dropped. It is also easy to show that second term in the figure brackets is infinitivally small in the case of paraxial motion resulting in non-relativistic motion in the co-moving frame:

$$\gamma_o^2\left(\frac{\beta_o}{\beta_z} - 1\right) = \frac{\gamma_o^2 \vec{\beta}_\perp^2}{\beta_z(\beta_o + \beta_z)} + \frac{\gamma - \gamma_o}{\gamma\beta_z(\beta_o + \beta_z)} \sim \frac{\vec{\beta}_{cm\perp}^2}{2} + \frac{\delta\gamma}{2\gamma} \ll 1. \tag{E4}$$

Specifically, $c\gamma_o\vec{\beta}_\perp$ is the transverse velocity in the co-moving frame and $\frac{\delta\gamma}{\gamma}$ is the relative energy deviation in the beam. Both of these values assumed to be infinitesimally small. As the result, the perturbation of the Hamiltonian is reduced to as simple:

$$\delta h^* = \frac{1}{\beta_o \gamma_o^2} \frac{e}{c} \delta\varphi = \frac{4\pi e^2}{c} \int \frac{\tilde{\rho}_{\vec{k}}}{\gamma_o^2 \beta_o^2 \vec{k}_\perp^2 + k_z^2} e^{i\vec{k}\vec{q}} \frac{d\vec{k}^3}{(2\pi)^3};$$
$$\tilde{\rho}_{\vec{k}} = \int e^{-i\vec{k}\vec{q}} \tilde{f}(q,p,s) dp^3 dq^3.$$
(E5)